\documentclass[%
 reprint,
nofootinbib,
 amsmath,amssymb,
 aps
]{revtex4-1}

\usepackage{graphicx}
\usepackage{xcolor}
\usepackage{dcolumn}
\usepackage{bm,bbold}
\usepackage{hyperref}

\def\SU #1{\texorpdfstring{$SU(#1)$}{SU(#1)}}

\newcommand{\units}[1]{\ensuremath{\,\mathrm{#1}}}
\newcommand{\eV}{\units{eV}}

\newcommand{\GeV}{\units{GeV}}

\newcommand{\TenFermion}{\Psi_{10}}
\newcommand{\FiveFermion}{\Psi_{\overline{5}}}
\newcommand{\OneFermion}{\Psi_{1}}
\newcommand{\FiveHiggs}{\Phi_5}
\newcommand{\FiveHiggsP}{{\Phi'}_{\!\!5}}
\newcommand{\TenHiggs}{\Phi_{10}}
\newcommand{\Yfive}{Y_{\overline{5}}}
\newcommand{\Yten}{Y_{10}}
\newcommand{\Yone}{Y_1}
\newcommand{\YtenP}{Y_{10}'}
\newcommand{\YfiveP}{Y_{\overline{5}}'}
\newcommand{\YoneP}{Y_1'}

\DeclareMathOperator{\diag}{diag}
\DeclareMathOperator{\im}{Im}
\DeclareMathOperator{\Tr}{Tr}

\def\N{\ensuremath{\mathcal{N}}}
\begin{document}
\title{Leptogenesis in the minimal flipped \SU5 unification}

\author{Renato Fonseca}\email{renatofonseca@ugr.es}
 \affiliation{Departamento de Física Teórica y del Cosmos, Universidad de Granada, Campus de Fuentenueva, E-18071 Granada, Spain}

\author{Michal Malinsk\'y}\email{malinsky@ipnp.mff.cuni.cz}
 \affiliation{Institute of Particle and Nuclear Physics,
  Faculty of Mathematics and Physics,
  Charles University in Prague, V Hole\v{s}ovi\v{c}k\'ach 2,
  180 00 Praha 8, Czech Republic}
 
\author{V\'aclav Mi\v{r}\'atsk\'y}\email{mirats.v@email.cz}
\affiliation{Institute of Particle and Nuclear Physics,
  Faculty of Mathematics and Physics,
  Charles University in Prague, V Hole\v{s}ovi\v{c}k\'ach 2,
  180 00 Praha 8, Czech Republic}

\author{Martin Zdr\'ahal}\email{zdrahal@ipnp.mff.cuni.cz}
 \affiliation{Institute of Particle and Nuclear Physics,
  Faculty of Mathematics and Physics,
  Charles University in Prague, V Hole\v{s}ovi\v{c}k\'ach 2,
  180 00 Praha 8, Czech Republic}

\date{\today}

\begin{abstract}
We study the prospects of thermal leptogenesis in the framework of the minimal flipped \SU5 unified model in which the RH neutrino mass scale emerges as a two-loop effect. Despite its strong suppression with respect to the unification scale which tends to disfavor leptogenesis in the standard Davidson-Ibarra regime (and a notoriously large washout of the $N_1$-generated asymmetry owing to a top-like Yukawa entry in the Dirac neutrino mass matrix) the desired $\eta_B\sim 6\times 10^{-10}$ can still be attained in several parts of the parameter space exhibiting interesting baryon and lepton number violation phenomenology. Remarkably enough, in all these regions the mass of the lightest LH neutrino is so low that it yields $m_\beta \lesssim 0.03\eV$ for the effective neutrino mass measured in beta-decay, i.e., an order of magnitude below the design sensitivity limit of the KATRIN experiment. This makes the model potentially testable in the near future.     
\end{abstract}

\maketitle

\section{Introduction\label{sect:introduction}}
The minimal flipped-\SU5 of~\cite{Rodriguez:2013rma} is a very attractive beyond-Standard Model framework, which, due to its simplicity, may be viewed as one of the most compact models of perturbative baryon and lepton number violation (BLNV) on the market. Its predictive power derives mainly from the very tight correlation between the structure of the BLNV currents governing processes like proton decay and the structures underpinning the neutrino sector. 
The key feature here is the fact that the RH Majorana neutrino masses
emerge only as a radiative effect at two loops, cf.~\cite{Witten:1979nr,Leontaris:1991mq,Rodriguez:2013rma}. 

This, among other things, sets the seesaw scale rather low compared to other unified models with the tendency to prefer a relatively large light-neutrino mass scale. As an implication, however, there is a potential challenge to a successful thermal leptogenesis in this scenario as, e.g., the basic Davidson-Ibarra~\cite{Davidson:2002qv} limit may be difficult to conform. 
On top of that, the $N_1$-asymmetry washout tends to be enhanced thanks to the specifics of the neutrino Yukawa pattern (with large $\propto y_t$ elements within) imposed by the enhanced gauge symmetry.
Hence, leptogenesis in the current scenario calls for a standalone analysis, which we provide here.

Needless to say, due to the relatively large number of parameters at play and the complexity of the relevant Boltzmann's equations (which, for reasonable precision, should include flavor/decoherence effects, etc., cf.~\cite{Blanchet:2011xq}), one has to primarily resort to a numerical approach. To this end, we shall be using the ULYSSES package~\cite{Granelli:2020pim} augmented with several analytic insights derived mainly from the special shape of the underlying seesaw formula.

The results of the analysis are quite remarkable -- a successful thermal leptogenesis in the minimal flipped SU(5) is not only possible, but it has a strong discrimination power concerning the favorable domains in the parameter space of the model. The main feature here is a strict upper limit on the mass of the lightest active neutrino, well below the sensitivity band of the current $\beta$-decay experiments such as KATRIN ~\cite{KATRIN:2001ttj,KATRIN:2022ayy}, which makes the model testable in the near future.  

The study is organized as follows: In Section~II, we describe the salient features of the minimal flipped SU(5) model, focusing in particular on the analytic structure of the seesaw formula and the implied multiplication rule for the physical masses of light and heavy neutrinos which happen to be strongly correlated to the up-type quark masses. In Section~III, we recapitulate the basics of the lepton asymmetry generation from the out-of-equilibrium decays of RH neutrinos in the early Universe (and its subsequent transformation into the baryon asymmetry), paying particular attention to the structure of the washout factors in the current model. This information serves as an input to Section~IV, where boundaries of viable domains in the model's parameter space are identified semi-analytically and then explored in full detail by numerical methods in Section~V. The impact of the constraints from leptogenesis on the structure of the BLNV currents and the implications for selected proton decay branching ratios are also discussed there.
Most of the technical details are deferred to a set of Appendices. 

In all this, we shall predominantly focus on the normal neutrino hierarchy case (following the global-fit preference, cf.~\cite{DeSalas:2018rby,Jimenez:2022dkn} but also \cite{Gariazzo:2022ahe}) and only briefly comment on the (mostly small) qualitative changes the results would undergo if the neutrino mass hierarchy is inverted. 

\section{The minimal flipped \SU5 model}
The basic structure of the minimal flipped \SU5 model under consideration is described in great detail in the previous studies~\cite{Rodriguez:2013rma,Harries:2018tld}. Hence, in what follows we shall focus only on its key features that are relevant to the discussion below. An interested reader can find further details in Appendices~ \ref{app:definitions} and~\ref{app:global_features}.
\subsection{The model}
\begin{figure}[t] \includegraphics[width=4.4cm]{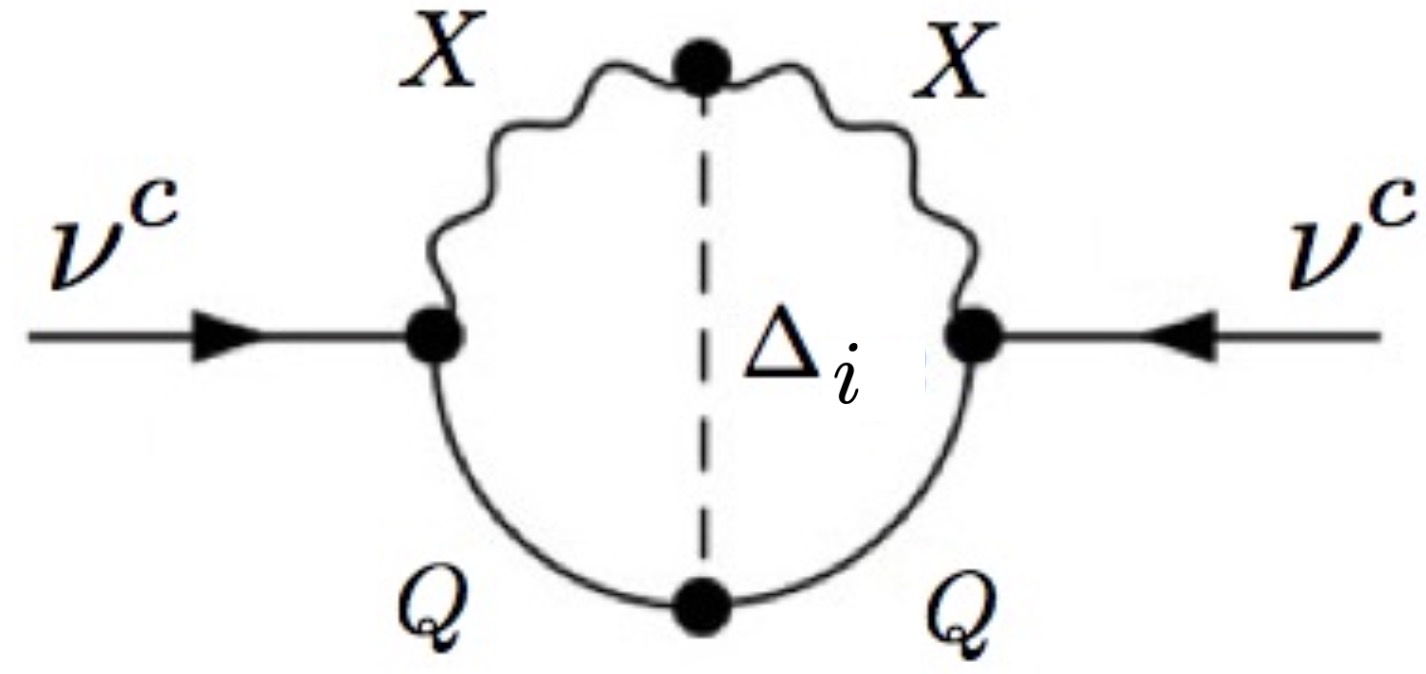}\quad
\includegraphics[width=4.4cm]{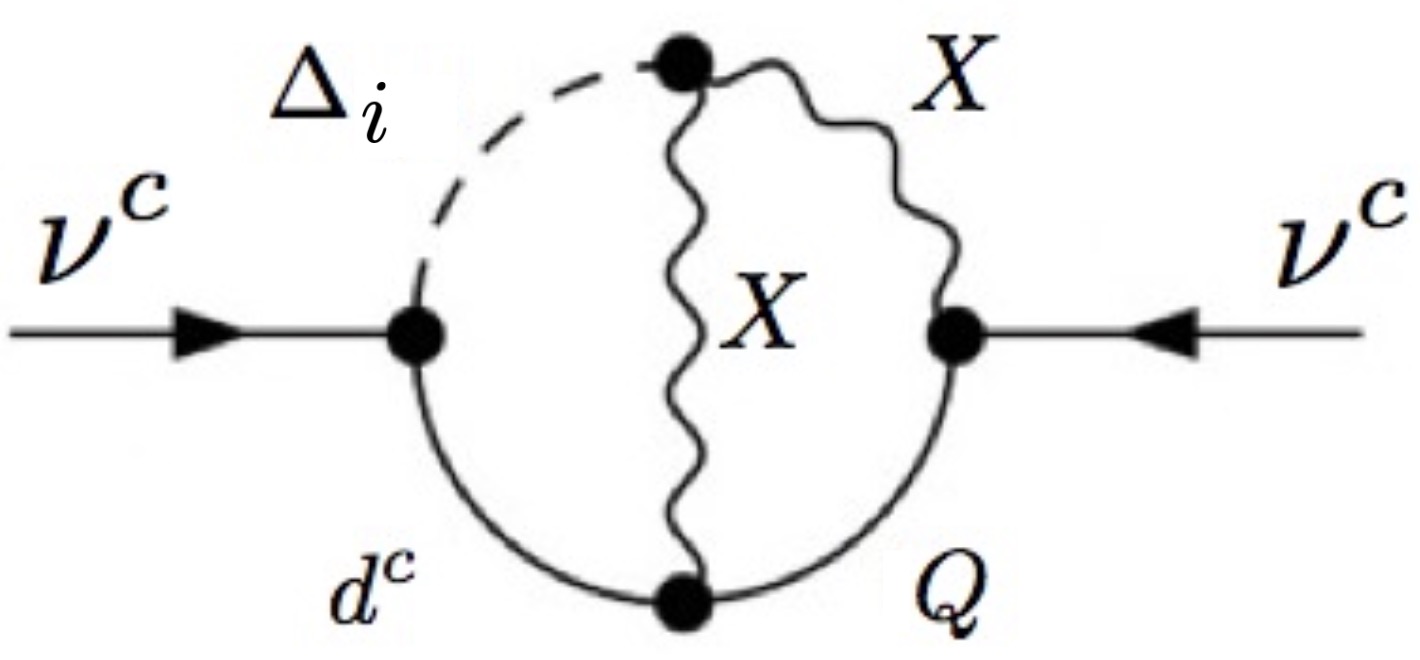}\caption{\label{fig:graphs}The two-loop diagrams providing the dominant contribution to the Majorana mass for the RH neutrinos in the minimal flipped $SU(5)$ model under consideration. The three color triplets $\Delta_i$ transforming as $(3,1,-\tfrac{1}{3})$ under the SM gauge group are coupled to the fermionic currents through their components in $\Phi_5$ and $\Phi'_5$, each coming with its own Yukawa matrix ($Y_{10}$ and $Y_{10}'$, respectively), cf.\ Eqs.~\eqref{eq:two-five-yukawa-sector} and \eqref{eq:neutrino formula}. The $X$ symbol denotes the $(3,\overline{2},+\tfrac{1}{6})$ massive gauge boson.  
    }
\end{figure}

Each generation of the Standard Model (SM) fermions (augmented by an extra right-handed neutrino, RHN) is accommodated in just three irreducible representations of the $SU(5)\times U(1)_X$ gauge group transforming as
\begin{equation}
\FiveFermion \equiv (\overline{5}, -3), \;\;
\TenFermion \equiv (10, +1), \;\;
\OneFermion \equiv (1, +5),
\end{equation}
with the extra $X$ charge connected to the SM hypercharge via
\begin{equation}
    Y=\frac{1}{5}(X-T_{24})\,,
\end{equation}
where $T_{24}$ is the usual fourth Cartan of the $SU(5)$, cf.~\cite{Rodriguez:2013rma}.
These matter fields interact with the scalar sector (consisting of one $\TenHiggs \equiv (10,+1)$ and two copies of $(5,-2)$ denoted hereafter  by $\Phi_5$ and $\Phi'_5$) via the Yukawa lagrangian of the form
\begin{align}
  \mathcal{L} &\ni \Yten \TenFermion \TenFermion \FiveHiggs
    + \YtenP \TenFermion \TenFermion \FiveHiggsP \nonumber \\
  & + \Yfive \TenFermion \FiveFermion \FiveHiggs^*
    + \YfiveP \TenFermion \FiveFermion \FiveHiggsP^* \nonumber \\
  &+ \Yone \FiveFermion \OneFermion \FiveHiggs
    + \YoneP \FiveFermion \OneFermion \FiveHiggsP + H.c.
    \label{eq:two-five-yukawa-sector}
\end{align}
with $\Yten$ and $\YtenP$ representing 3$\times$3 complex symmetric matrices, and $\Yfive$, $\YfiveP$, $\Yone$, and $\YoneP$ standing for general 3$\times$3 complex matrices in the matter generation space. The reason for having two scalar pentuplets rather than one has to do with the need to smear the overly strong correlation\footnote{In principle, both $Y_{10}$ and $Y'_{10}$ contribute to $M_d$ and $M_\nu^M$. However, their relative weights in these combinations are generally different and one can always go into the basis in the $\Phi_5$-$\Phi'_5$ space where one of them is just $M_d$ and the other one is left unconstrained. cf. Appendix~\ref{app:global_features}. 
} between the down quark mass matrix $M_d$ and the RH neutrino Majorana mass matrix $M_\nu^M$, cf.~\cite{Rodriguez:2013rma}. The latter is generated radiatively via the two-loop diagrams in Fig.~\ref{fig:graphs} and reads\footnote{Note that in \cite{Harries:2018tld}, there is a typo in expressions (10), (18), and (24), where a factor of two (connected with two possible contractions of the vector fields) is missing.} 
\begin{align}
    \label{eq:neutrino formula}
		M_\nu^M=\frac{6g_5^4}{(4\pi)^4}V_G&\sum_{i=1}^{3}\Big(
		8Y_{10}[U_\Delta]_{i1}[U_\Delta^*]_{i2}\\
	&\quad\,+8Y_{10}'[U_\Delta]_{i1}[U_\Delta^*]_{i3}
		\Big)I_{3}\Big(\frac{m^2_{\Delta_i}}{m_X^2}\Big),\nonumber
\end{align}
where $g_5\approx 0.5$ (see~\cite{Rodriguez:2013rma}) is the unified \SU5 gauge coupling, $V_G$ is the vacuum expectation value of the $\Phi_{10}$ scalar responsible for, e.g., the  generation of the masses of the gauge bosons (such as $X$ in Fig.~\ref{fig:graphs} with 
$m_X^2=g_5^2V_G^2/2$) and $i$ runs over the three different coloured triplets $\Delta_{1,2,3}$, whose mass matrix is diagonalized by a unitary transformation $U_\Delta$, cf.\ Appendix~B of \cite{Harries:2018tld}. Finally, $I_3$ is a loop function (monotonous with a range of $(-3,3)$) depending on the indicated ratio of heavy triplet ($\Delta$) and $X$ masses; an interested reader can see its behavior in Fig.~2 of \cite{Harries:2018tld}.

\subsection{The seesaw\label{sect:seesaw}}
Following the definitions and conventions fixed in Appendix~\ref{app:seesaw_anatomy} the light-neutrino Majorana mass matrix $m_{LL}$ can be written like
\begin{equation}\label{original mLL}
	m_{LL}=-M_\nu^D\left(M_\nu^M\right)^{-1}\left(M_\nu^D\right)^{T}
\end{equation}
where $M_\nu^D$ is the relevant Dirac neutrino mass matrix; due to its symmetric shape it can be diagonalized by means of a unitary transformation
\begin{equation}\label{diagonalization mLL}
	m_{LL}=U_\nu^\dagger D_\nu U_\nu^*\,.
\end{equation}
Note that the two formulae above apply at the seesaw scale $M_{BL}\sim |M_\nu^M|$ and, hence, the eigenvalues of $m_{LL}$ contained in $D_\nu$ are not exactly the low-energy light neutrino masses (although they are directly related to them, cf.\ Appendix~\ref{app:inputs}). Similarly, the key relation between $M_\nu^D$ and $M_u$ available in the flipped \SU5 framework under consideration, namely, 
\begin{equation} \label{MnuD_Mu_correspondence}
	M_\nu^D=M_u^T,
\end{equation}
cf.~\cite{Rodriguez:2013rma}, holds at the unification scale $M_5\sim 10^{16}\GeV$ and, thus, relates $M_\nu^D$ to the running quark masses evolved up to $M_5$. 
All these nuances shall be taken into account in the numerical analysis of Section~\ref{sect:Results}.

In what follows, it is convenient to extract the RH neutrino mass matrix from Eq.~(\ref{original mLL}) and write it as
\begin{equation}\label{seesaw_diagonal_form}
	M_\nu^M \equiv \tilde{U}^T D_\nu^M\tilde{U}=- \hat{D}_u U_\nu^T D_\nu^{-1} U_\nu \hat{D}_u, 
\end{equation}
where $U_\nu$ and $D_\nu$ are seesaw-scale quantities, and 
\begin{equation}
\hat{D}_u\equiv {\rm diag}\{\hat{m}_u,\hat{m}_c,\hat{m}_t\} \end{equation} is a shorthand for the eigenvalues of the up-type quark mass matrix $M_u$ evolved from $M_5$ to $M_{BL}$ by means of the RGEs for\footnote{Here we strongly benefit from the fact that one can, in principle,
switch to the basis in which $M_\nu^D= M_u^T$ (that is valid at $M_5$) are
both diagonal matrices, cf.\ Appendix B of Ref.~\cite{Rodriguez:2013rma}, and that any
(tiny) non-diagonality of $M_\nu^D$ generated by pulling it from $M_5$
down to $M_{BL}$ may be subsumed into redefinitions of $U_\nu$ and $\tilde{U}$.} $M_\nu^D$. 
The eigenvalues of $M_u$ at $M_5$ are connected to the low-energy data as described in Appendix~\ref{app:inputs}; the key observation is that their ratios are very close to those of the low-energy quark masses, cf.~\eqref{eq:upquarkratios}, which makes $\hat{D}_u$ strongly hierarchical.    

Note that Eq.~\eqref{seesaw_diagonal_form} yields a nice product rule for the quantities of our main interest, namely,
\begin{equation}\label{eq:product_rule}
\hat m_1 \hat m_2 \hat m_3 M_{1}M_{2}M_{3} = \hat m_u^{2}\hat m_c^{2}\hat m_t^{2}
\end{equation}
where $\hat m_i$ correspond to the light neutrino masses evolved to the seesaw scale. This evolution effect will be, in the first approximation, accounted for by a simple multiplication of their low-scale values ($m_i$) by a common $r$-factor, i.e., 
\begin{equation}
 \hat m_i= r\,m_i\,,  
\end{equation}
where $r\,\sim 1.32$ (cf. Ref.~\cite{Antusch2003,Xing2008} and Appendix~\ref{app:inputs}).

With all this at hand, one can write the Pontecorvo-Maki-Nakagawa-Sakata leptonic mixing matrix $V_{PMNS}$ (in the `raw' form, i.e., without the three unphysical phases stripped off, cf. Appendices~\ref{app:seesaw_anatomy} and~\ref{App:Basis change}) as 
\begin{equation}\label{PMNS}
	V_{PMNS}=U_\ell^L U_\nu^\dagger,
\end{equation}
where the $U_\ell^L$ factor comes from the bi-unitary diagonalization of the charged lepton mass matrix
\begin{equation}\label{lepton mass matrix}
	M_\ell=(U_\ell^L)^\dagger D_\ell U_\ell^R.
\end{equation}
Note that there is a rationale behind the seemingly different conventions in the definitions of the $U_\nu$ and $\tilde U$ diagonalization matrices in Eqs.~\eqref{diagonalization mLL} and~\eqref{seesaw_diagonal_form}, (namely, the extra complex conjugation in the former); an interested reader can find more details in Appendix~\ref{app:seesaw_anatomy}. 

\section{Leptogenesis in flipped $SU(5)$}
The presence of relatively light RH neutrinos (RHNs) subject to multiple additional constraints\footnote{
It is remarkable that in spite of starting with 6 independent Yukawa matrices in the defining Lagrangian~\eqref{eq:two-five-yukawa-sector} the current model turns out to be way more predictive than the minimal SM extension with RH neutrinos which feature only 5 such flavour-defining structures. This has to do with the fact that in flipped $SU(5)$ the charged leptons alone ``consume" two of these Yukawas and the identity~\eqref{MnuD_Mu_correspondence} holds irrespective of the actual number of scalar pentuplets employed.  } on the shape of their Yukawa interactions~\eqref{diagonalization mLL} and, in particular, their spectrum~\eqref{eq:product_rule}, calls for a dedicated analysis of the leptonic asymmetry generated in their out-of-equilibrium decays. In what follows, we shall first provide a basic account of thermal leptogenesis in this context and then apply the formalism to the situation of our interest.
\subsection{RHN-driven thermal leptogenesis}
The quantity of central importance in this type of scenarios is the leptonic CP asymmetry of the decays of the $i$'th heavy neutrino mass eigenstate $N_i$ into the SM leptons $L$ (and antileptons $\overline{L}$), namely 
\begin{equation}\label{CP_definition}
	\epsilon^i_{CP} \equiv\frac{\Gamma(N_i\to\phi L)-\Gamma(N_i\to\phi^\dagger\overline{L})}{\Gamma(N_i\to\phi L)+\Gamma(N_i\to\phi^\dagger\overline{L})}\,.
\end{equation}
As it turns out, the flavour structure of the lepton decay states gives rise to numerous effects, which can affect the whole process of generation of the CP asymmetry significantly \cite{Blanchet:2011xq}. To briefly demonstrate this, let $\vert i \rangle$ denote the flavor superposition states of leptons originating from the aforementioned decays, and let $\vert \alpha \rangle$ denote the lepton flavour eigenstates. The decay states can be expressed as
\begin{equation} \label{eq:lepton_c}
    \vert i \rangle = c_{i\alpha} \vert \alpha \rangle
\end{equation}
and analogicaly for the antilepton case, one can write
\begin{equation} \label{eq:antilepton_c}
    \vert \bar{i} \rangle = \bar{c}_{i\bar{\alpha}} \vert \bar{\alpha} \rangle\,.
\end{equation}
Even though the SM gauge interactions preserve the CP asymmetry, they still play an important role. For specifity, consider the leptons produced by the decay of $N_2$ as an example. Such leptons can naturally transform back to $N_2$ via inverse decays, but potentially also to $N_1$. However, the electroweak gauge interactions of the leptons with the thermal bath lead to so called decoherence effects. These interactions are flavor-diagonal, so the flavour superposition $\vert 2 \rangle$, after an interaction with the electroweak bosons, transforms into a pure flavor state. If this phenomenon can be neglected, the evolution of the state is said to be coherent. On the contrary, if this effect is strong, meaning all decay states manage to interact with the bath through electroweak interactions before they decay back to RH neutrino, we refer to this scenario as complete decoherence. It is evident that describing the situation between these limiting cases requires quantum statistics approach. Therefore, instead of using classical Boltzmann equations, equation for the effective leptonic flavor density matrix, defined in e.g \cite{Blanchet:2011xq,Granelli:2020pim}, must be employed.

To underscore the relevance of the flavour effects, consider the scenario in which the decay states evolve coherently. If the $\vert 2 \rangle$ was perfectly orthogonal to the $\vert 1 \rangle$ state, it is evident that the products resulting from the decay of $N_2$ could not be washed out just by inverse decays of $N_1$. Equivalently, if $\vert 2 \rangle$ has a non-zero projection onto the subspace orthogonal to $\vert 1 \rangle$, a proportional fraction of the $N_2$ decay states is protected from the washout due to $N_1$. This phenomenon is called the projection effect. Clearly, the situation is different if   decoherence occurs, i.e., when the $N_i$ decay states  experience other interactions before inverse decays that make them undergo transitions into flavor states. Moreover, if one is within a strong washout regime and the RH neutrino mass spectrum is hierarchical, any asymmetry originating from $N_2$ decays is incapable of persisting. The sole exception would arise if one of the coefficients $c_{1\alpha}$ is zero, while the $c_{2\alpha}$ is non-zero.

Using the density matrix formalism, the evolution of the the effective leptonic flavor density matrix $N^{B-L}_{\alpha\beta}$ is governed by a matrix-like Boltzmann evolution equation in the schematic form  
\begin{equation}\label{Boltzmann_DE}
\frac{{\rm d} N^{B-L}_{\alpha\beta}}{{\rm d} z}=\sum_i\mathcal{D}^i\epsilon^i_{\alpha\beta}-\mathcal{W}_{\alpha\beta}-\mathcal{C}_{\alpha\beta},
\end{equation}
where $z=M_1/T$ is the cosmological parameter corresponding to the ratio between the temperature and the mass of $N_1$; the term $\mathcal{D}^i\epsilon^i_{\alpha\beta}$ is the production term of the asymmetry in the relevant flavor channels depending, among other things, on the excess of the individual $N_i$'s over their equilibrium number densities, and $\mathcal{W}_{\alpha\beta}$ and $\mathcal{C}_{\alpha\beta}$ are the washout and decoherence terms, respectively, both depending on $N^{B-L}_{\alpha\beta}$ at the given moment. For a detailed derivation and explicit formulas of the particular terms, see \citep{Blanchet:2011xq}.

The resulting total baryon asymmetry of the Universe written in terms of the notorious baryon-to-photon number density ratio 
\begin{equation}
	\eta_B=\frac{n_B-n_{\overline{B}}}{n_\gamma}
\end{equation}
is then theoretically calculated to be
\begin{equation} \label{eta_theor}
    \eta^{\rm th.}_B\approx 10^{-2}\times \Tr N_{f}^{B-L},
\end{equation}
where $N_{f}^{B-L}$ denotes the asymptotic ($z\to \infty$) value of $N^{B-L}$ determined from \eqref{Boltzmann_DE}, and the trace is taken over the flavor indices. This should eventually be compared to the observed value of 
\begin{equation}\label{etab_obs}
	\eta_B^{\rm obs.}=6.1\times 10^{-10}\,.
\end{equation}

\subsection{Leptogenesis in the minimal flipped $SU(5)$\label{sec:SU5leptogenesis}}
\subsubsection{General parametrization}
Prior to addressing the central question of this study, namely, whether the thermal leptogenesis picture can be supported in the current scenario (featuring constraints like~\eqref{MnuD_Mu_correspondence} and~\eqref{eq:product_rule}), it is necessary to identify the independent parameters of the model that determine the resulting baryon asymmetry \eqref{eta_theor}. In order to do so, one has to realize~\cite{Blanchet:2011xq} that the RHS of the equation \eqref{Boltzmann_DE} is only a function of the density matrix itself, the masses of $N_i$, and the Yukawa matrix $Y_{\nu}$, which comes from 
\begin{equation}
\mathcal{L} \ni  
Y_{\nu} \bar{L}N\phi + H.c.
\end{equation}
and mediates the decays \eqref{CP_definition}. However, a key feature of the flipped $SU(5)$ is that the Dirac mass spectrum is, in principle, known at the seesaw scale (as explained in Section~\ref{sect:seesaw}) and there is a strong constraint on the RH neutrino spectrum~\eqref{eq:product_rule}. It turns out that in this situation, instead of the traditional Casas-Ibarra parametrization~\cite{Casas:2001sr}, it is more convenient to employ the standard singular value decomposition, which yields\footnote{This expression is fully compatible with the Casas-Ibarra parametrization \cite{Casas:2001sr}; indeed,
$
	vY_\nu= i V_{PMNS} D_\nu^{1/2}\mathcal{R}(D_\nu^M)^{1/2}
$
for $\mathcal{R}$ written as
$
	\mathcal{R}=-i D_\nu^{-1/2} U_\nu  \hat{D}_u \tilde{U}^\dagger (D_\nu^M)^{-1/2}.
$
},
\begin{equation}\label{Yukawa_from_U}
	Y_\nu=\frac{1}{v}U_\ell^L \hat{D}_u \tilde{U}^\dagger=\frac{1}{v}V_{PMNS}U_\nu \hat{D}_u \tilde{U}^\dagger\,
\end{equation}
(see also Eq.\ \eqref{eq:YnuU1} in Appendix \ref{app:definitions}) with $v=174\GeV$ denoting the SM Higgs field vacuum expectation value. In writing~\eqref{Yukawa_from_U} we have implicitly adopted the convention defined by Eq.~\eqref{seesaw_diagonal_form}, i.e., switched to the basis in which the heavy neutrino and charged lepton mass matrices are diagonal, see definitions~\eqref{PMNS} and~\eqref{lepton mass matrix}.  
Note also that there is a strong correlation among the $\tilde{U}$ and the $U_{\nu}$ matrices (and the associated light and heavy neutrino spectra) provided by the seesaw formula \eqref{seesaw_diagonal_form} so given one side (typically $U_\nu$
along with the light neutrino spectrum) the other side is fully determined ($\tilde{U}$ and the heavy neutrino spectrum). 

Subsequently, with the experimental data from Appendix~\ref{app:inputs} at hand, the only remaining free parameters are the mass of the lightest neutrino $m_0$, the light neutrino Majorana phases $\alpha_{1,2}$ and the unitary matrix $U_{\nu}$. Since three phases on the RH side (RHS) of the $U_{\nu}$ matrix can be absorbed in the redefinition of phases of $\tilde{U}$ (which are irrelevant in the current context, see Appendix~\ref{App:Basis change}), there remains a total of 9 free parameters determining the resulting baryon asymmetry.
\subsubsection{Washout}
Another distinct feature of the model is the presence of top-like elements in the Yukawa matrix $Y_{\nu}$. Defining the washout parameter for the $i$'th heavy neutrino mass eigenstate $N_i$ as the ratio of its total decay width to the Hubble rate evaluated at $T=M_i$, one arrives at
\begin{align}\label{washout_ki}
    k_i\equiv\frac{\Gamma(N_i)}{H(T=M_i)} &= \frac{v^{2}}{m_* M_i} \left(Y_\nu^T Y_\nu^*\right)_{ii}\\
    &=\frac{|\tilde{U}_{i3}|^2 \hat m_t^2+|\tilde{U}_{i2}|^2 \hat m_c^2 +|\tilde{U}_{i1}|^2 \hat m_u^2}{m_* M_i}\,,
    \nonumber
\end{align}
where the equilibrium neutrino mass (cf.~\cite{Buchmuller:2004nz}) assumes
\begin{equation}
    m_* \approx 10^{-3} \eV.
\end{equation}
Note that if the $\hat m_t^2$-proportional term in~\eqref{washout_ki} is not suppressed, the associated washout factor tends to be way larger than 1 (of the order of $10^{16}{\rm GeV}/M_i$), indicating a strong washout regime for $N_i$. In such a case, the contribution of the decay of $N_i$ to the total lepton number asymmetry $N_{f}^{B-L}$ is bounded from above by \cite{Buchmuller:2004nz}
\begin{equation}\label{asymmetry_limit}
\left(N_{f}^{B-L}\right)_{i} \lesssim \frac{2\epsilon_{CP}^i}{k_i}\,,    
\end{equation} 
where the relevant CP asymmetry is
\begin{multline} \label{epsilon i from Y}
\epsilon_{CP}^i=\sum_\alpha \epsilon_{\alpha\alpha}^i\\ 
= -\frac{1}{8\pi} \frac{1}{[Y_\nu^T Y_\nu^*]_{ii}}\sum_j\im\left[\left(Y_\nu^T Y_\nu^*\right)_{ij}^2\right]f\left(\frac{M_i^2}{M_j^2}\right),
\end{multline}
provided
\begin{equation} \label{eq:CP_loop_function}
	f(x)=\sqrt{x}\left(-\frac{2-x}{1-x}+(1+x)\ln \frac{1+x}{x}\right).
\end{equation}
Such a strong washout is a clear challenge to the current scenario as it is not guaranteed that a sufficiently high CP asymmetry can be simultaneously attained to compensate for it. Indeed, Davidson and Ibarra~\cite{Davidson:2002qv} stipulate that for hierarchical heavy neutrino masses the CP asymmetry generated in the $N_1$ decays is  bounded from above by\footnote{This relation holds for the normal light neutrino hierarchy. For the inverted case one would have instead 
\begin{equation}\label{eq:DINH}
	\left|\varepsilon_{CP}^1\right|\lesssim
	\frac{3}{8\pi}\frac{M_1}{v^2} \frac{\Delta m_{31}^2+\Delta m_{21}^2}{m_2+m_3}\,r\,.
\end{equation}
} 
\begin{equation}\label{DavidsonIbarra}
	\left|\varepsilon_{CP}^1\right|\lesssim
	\frac{3}{8\pi}\frac{M_1}{v^2} \frac{|\Delta m_{31}^2|}{m_3+m_1}\,r\,
\end{equation}
and, hence, the CP asymmetry generated in the $N_1$ decay is typically large for large  $M_1$ (and low $m_{1})$. However, with radiatively generated RH neutrinos as in the current scenario such configurations can be problematic as they may, for instance, require large values of the associated couplings, with a potential conflict with perturbativity (cf. Appendix~\ref{app:perturbativity}) or, for non-negligible $m_1$, may be at odds with other model constraints such as~\eqref{eq:product_rule}.
Therefore, in order to attain large enough $\eta_B$, one may expect to be pushed to specific domains in the parameter space where the critical $|\tilde U_{i3}|$ weight factor associated with $\hat{m}_t$ in~\eqref{washout_ki} is suppressed, cf. Appendix~\ref{app:minimal_washout}. 
Another potentially viable strategy would be to consider a quasi-degenerate RH neutrino mass spectrum for which the  strict Davidson-Ibarra limit is relaxed and the CP asymmetry can be significantly enhanced\footnote{Note that this regime can be achieved  relatively easily even without resorting to the magic of resonant leptogenesis~\cite{Pilaftsis:2003gt} which, from the current scenario perspective, is very unnatural as there is no reason for the required delicate fine-tuning in the RHN spectrum.}.

As we shall see in Section \ref{sect:Results}, it is not so difficult to find points in the parameter space corresponding to the two regimes above for which the baryon asymmetry is consistent with the observed value \eqref{etab_obs} and the mass of the lightest neutrino mass eigenstate (to be from now on denoted by $m_0$) is large enough to avoid all perturbativity constraints trivially (cf. Appendices~\ref{app:perturbativity} and~\ref{app:perturbativity_further}). Hence, the issue of perturbativity will not be the primary concern here.

\subsubsection{Running parameters}
Let us finish this section by pointing out that all the factors on the RHS of Eq.~\eqref{Yukawa_from_U} contain seesaw-scale evaluated quantities. Concerning the shape of $V_{PMNS}$, it is safe to use its low-energy form therein for at least two reasons: First, in most of the leptogenesis-relevant structures (like CP asymmetries and washout factors), $Y_\nu$ enters as $Y_\nu^\dagger Y_\nu$, where $V_{PMNS}$ plays no role. Second, the running effects in the leptonic mixing parameters between the $M_Z$ and the $M_1$ scale are negligible in situations that we shall be dealing with in Sec.~\ref{sect:Results}. The only quantities in which the running may, in principle, play a role are the flavor parameters \eqref{eq:lepton_c} and \eqref{eq:antilepton_c}. Indeed, in the one-loop approximation, one has \cite{Blanchet:2011xq}
\begin{equation}
    \label{eq:cialphaasYukawa}
    c_{i\alpha} = \bar{c}^{\star}_{i\bar{\alpha}} = \frac{Y_{\alpha i}}{\sqrt{(Y^{\dagger}Y})_{ii}}\,,
\end{equation}
where the dependence on the $V_{PMNS}$ is evident. 

However, rather than the actual numerical values of these coefficients, it is namely their qualitative features (such as vanishing of some of them or very special mutual orientation of the entire decay states vectors) that govern the resulting leptogenesis yields. In this sense, the extra freedom in the unknown Majorana phases sticking out of the RHS of $V_{PMNS}$ in formula~\eqref{Yukawa_from_U} ensures that any change in the specific shape of $V_{PMNS}$ that can be reasonably expected due to running effects should not matter much when it comes to attaining a specific regime for $c_{i\alpha}$'s. 
The same reasoning can also be used in arguing that any potential dependency of the resulting $\eta_B$ on the the Dirac CP phase $\delta$ in $V_{PMNS}$ is likely to be screened in the freedom to adjust these and other unknown parameters; hence, in what follows, we shall just set $\delta$ to its current central value of $1.08\,\pi$ (cf.~\cite{deSalas:2020pgw}) rather than taking it as another free parameter.      

\section{Parameter space topography\label{sect:parameter_space}}
Before approaching the numerical analysis it is convenient to make several analytical observations concerning the relevant input parameters and, hence, facilitate the mapping of the the most promising (from the leptogenesis perspective) regions of the parameter space\footnote{Let us reiterate here that the detailed analysis of Section~\ref{sect:Results} will focus on the normal light neutrino mass ordering (as suggested by the latest global fits, cf.~\cite{DeSalas:2018rby,Jimenez:2022dkn}); the differences in the main results for the inverse hierarchy case will be briefly accounted for in Section~\ref{sect:inversehierarchy}.
}.

First, the three RHN masses obey the  rule~\eqref{eq:product_rule} and, hence, for a fixed light neutrino spectrum (parametrized by the mass of the lightest mass eigenstate $m_0$), their product is essentially fixed.  
The relevant parameter-space domains in the $M_i$-$M_j$ log-log plots then attain simple triangular shapes, cf.\ Section~\ref{sect:Results}.        

Second, a region of the parameter space where a large enough $\eta_B$ is attained can be relatively easily identified within the domain where the Davidson-Ibarra inequality~\eqref{eq:DINH} holds at least approximately. The point is that taking it as a lower limit on $M_1$ (rather than the traditional upper limit on $|\epsilon_{CP}^1|$), one has\footnote{Note that the numerical lower bound we take on the LHS of Eq.~\eqref{eq:M1limitfromDI} is extremely conservative as it essentially ignores the extra suppression factors encountered upon translating the leptonic asymmetry into the baryonic one, see Eqs.~\eqref{eta_theor} and~\eqref{asymmetry_limit}; in this sense, the range of validity of the  bound~\eqref{eq:M1bothsaturated} is larger than one may na\"\i vely expect.}   
\begin{equation}\label{eq:M1limitfromDI}
    10^{-9}\lesssim\frac{2|\epsilon_{CP}|}{k_1}\lesssim \frac{3}{4\pi} \frac{m_* M_1^2}{v^2 \hat m_u^2} \frac{\Delta m_{31}^2}{m_0+m_3}\,r,
\end{equation}
where the first inequality reflects the need for a large enough $\eta_B$; in the latter, the lower limit on the washout factor  
\begin{equation}\label{eq:k1min}
k_1\gtrsim\frac{\hat m_u^2}{m_* M_1}   
\end{equation} 
stemming from Eq.~\eqref{washout_ki} has been utilized. Note that the bound~\eqref{eq:k1min} is saturated for 
\begin{equation}\label{eq:k1saturation}
|\tilde U_{1i}|=\{1,0,0\}\,.   
\end{equation}
Assuming for the moment that this saturation occurs in the same domain where~\eqref{eq:M1limitfromDI} applies (which, obviously, does not need to be the case in general), one obtains two (perhaps somewhat na\"{\i}ve but still useful) estimates for $M_1$ and $\hat{m}_0$ (where the latter denotes the mass of the lightest light neutrinos evolved to the seesaw scale). As for the former, relations (\ref{eq:M1limitfromDI}) and \eqref{eq:k1min} yield (see also Appendix~\ref{app:inputs})
\begin{equation}\label{eq:M1bothsaturated}
M_1\gtrsim 10^7\GeV\,,  
\end{equation}
which, combined with the expression for the 11 element of the RHN mass matrix~\eqref{seesaw_diagonal_form} 
\begin{equation}\label{eq:M11diag}  |(M_\nu^M)_{11}|=|M_1|=\hat m_u^2 |\sum_i \hat m_i^{-1}(U_\nu)_{i1}^2|\,, 
\end{equation}
(where the first equality follows from~\eqref{eq:k1saturation}) leads to
$   
   {10^7\GeV}/{\hat m_u^2} \lesssim \hat m_0^{-1}\,,
$
and, eventually, to\begin{equation}\label{eq:limit}
    \hat m_0\lesssim 10^{-4.6}\eV.
\end{equation}
At one hand, an upper limit on the mass of the lightest active neutrino like this is phenomenologically interesting as it cuts deep into the  parameter space inaccessible to the current terrestrial neutrino-mass-oriented experimental searches such as KATRIN. 
On the other hand, inequality~\eqref{eq:limit} is by far not a hard limit on $\hat m_0$, as there were several assumptions made upon its derivation and, as such, it should be taken as a mere indication\footnote{On the other hand, the smaller $\hat m_0$ one takes, i.e., the further away from the limit quoted in Eq.~\eqref{eq:limit} one goes, the more room there is to depart from the minimal washout domain (defined by Eq.~\eqref{eq:k1saturation}) whilst maintaining large enough $\eta_B$ and, hence, the easier to find suitable points.} of where to look for a general and robust upper bound .  

With this information at hand, the general strategy of the numerical analysis of the next Section can be drawn: As we are mainly interested in mapping the boundaries of the parameter-space domain(s) supporting thermal leptogenesis in the current scenario, we shall be predominantly looking at the situation {\em above} the limit (\ref{eq:limit}). To this end, we shall start with a thorough numerical scan for $\hat m_0\sim 10^{-3}\eV$ and then, if possible, try to push $\hat m_0$ even higher. 

Concerning the numerical boundaries to be imposed onto the RHN masses, we shall generally respect the na\"\i ve limit~\eqref{eq:M1bothsaturated} because only in such a case the regimes in which $N_2$ (and/or $N_3$)-generated asymmetries may survive the $N_1$-induced washout may be attainable; indeed, for $M_1<10^7$~GeV, decoherence would be overly efficient for large enough $\eta_B$ to be produced in these scenarios.     

\section{Numerical analysis and results\label{sect:Results}}
Let us finally approach the numerical analysis of the leptogenesis yields which is performed along the following lines: First, for a fixed value of $\hat{m}_{0}$ we pseudo-randomly\footnote{Note that an efficient generation of points with sufficiently high masses $M_{1}$, for which one can expect the highest resulting baryon asymmetry, is by no means trivial. The method we have employed for that sake is discussed in detail in Appendix~\ref{app:generator}.} generate a sample of $U_{\nu}$ matrices \eqref{diagonalization mLL} and the low-energy Majorana phases (i.e., those affecting the flavor projections~\eqref{eq:lepton_c} and \eqref{eq:cialphaasYukawa} through the PMNS matrix in formula~\eqref{Yukawa_from_U}). With this at hand, the mass spectrum of the RHNs is calculated via the seesaw formula \eqref{seesaw_diagonal_form}. This, together with the fully determined Yukawa matrix $Y_{\nu}$~\eqref{Yukawa_from_U}, is then fed into the ULYSSES package~\cite{Granelli:2020pim} designed for solving the relevant Boltzmann's equations and evaluating the final baryon asymmetry for each such point.
To this end, let us reiterate that for the sake of this study we did not entertain the highly fine-tuned scenarios featuring resonant enhancement of $\eta_B$; hence, technically, all our calculations have been performed within the {\tt 3DME} model available therein. Finally, we illustrate the impact of the  leptogenesis constraints (especially those obtained for the shape of the $U_\nu$ matrix) on proton decay partial widths for a sample of points corresponding to one specific value of $\hat m_0$. 

\subsection {Upper limit on $\hat m_0$ from leptogenesis\label{sec:m0upperlimit}}
\begin{figure}[t!]
	\centering
	\mbox{
\includegraphics[height=8.5cm,width=8cm]{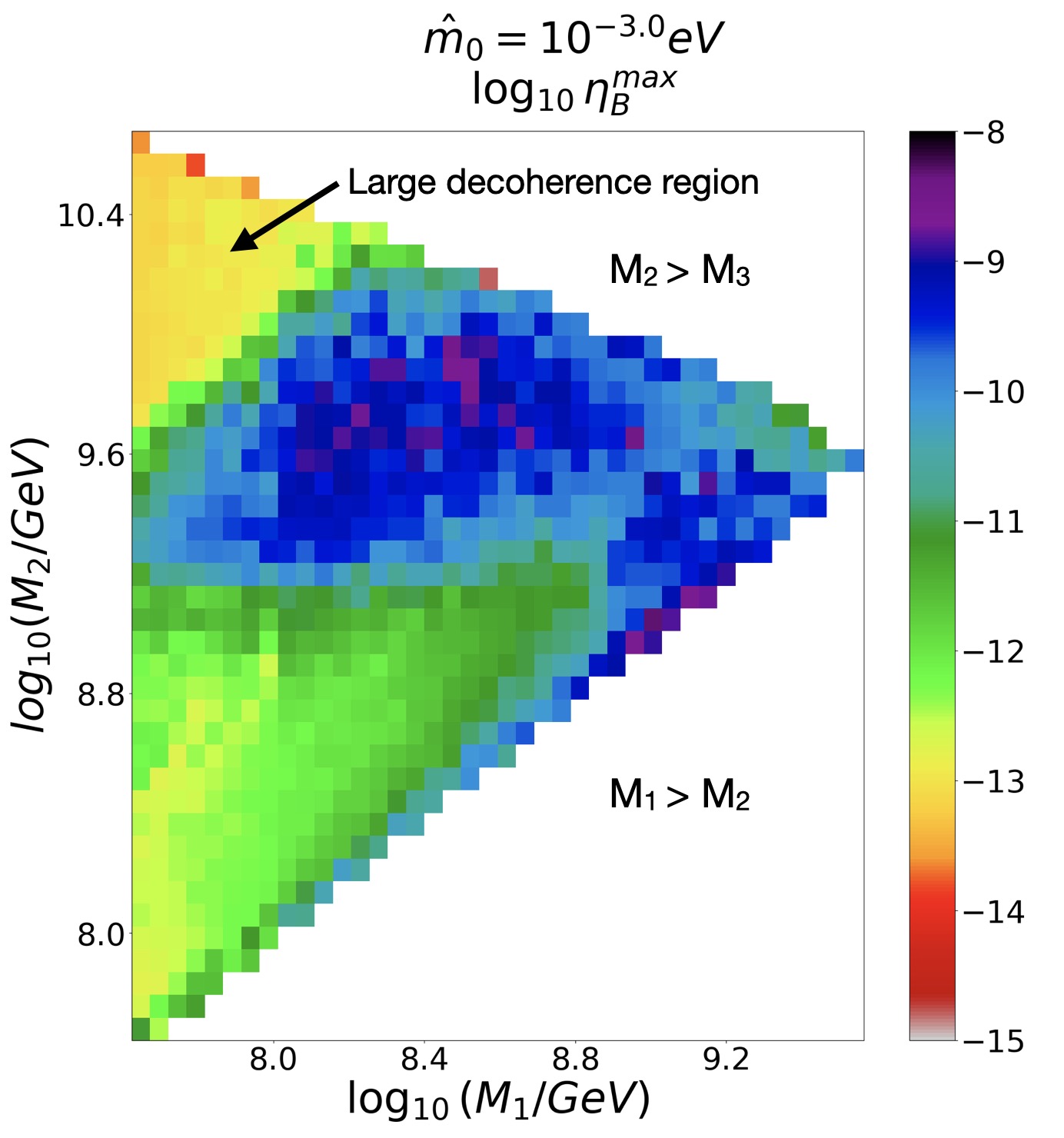}
    }
	\caption{
 The maximum attainable $\eta_B$ for different configurations of the heavy neutrino masses subject to the constraint~\eqref{eq:product_rule} with $\hat m_0=10^{-3}\eV$   
  (and normal neutrino hierarchy), cf. Section~\ref{sect:Results}. Note that for such a hierarchical light neutrino spectrum, $\eta_B\sim 6\times 10^{-10}$ is easy to achieve over a significant portion of the plot. For the analytic interpretation of the main features visible therein, see the discussion in Section~\ref{sec:m0upperlimit}. The limited resolution (granularity) of the plot is due to the complexity of the scan over the 8 free parameters requiring point-by-point integration of the Boltzmann equations. Note that about 15 millions of points had to be generated~(see Appendix~\ref{app:generator}) and processed in order to get this plot. 
 }
	\label{fig:maximum baryon asymmetry 1}
\end{figure}

In Figure \ref{fig:maximum baryon asymmetry 1}, we display the maxima of $\eta_{B}$ attained for different shapes of the right-handed neutrino spectra for the aforementioned $\hat m_{0}=10^{-3}\eV$. Note that the available domain for $M_{1}$ and $M_{2}$ does, indeed, resemble a triangle in the log-log plot as anticipated in Section~\ref{sect:parameter_space}, cf.\ also Eq.~\eqref{eq:product_rule}.
One can immediately see that, in this case, it is possible to get $\eta_B$ perfectly compatible with observation. On the other hand, large enough $\eta_B$ is not available everywhere in the $M_1$-$M_2$ chart and can be attained only in several areas therein (albeit with somewhat fuzzy boundaries). 
These features are even better seen in an equivalent plot where $\hat m_0$ has been increased to as much as\footnote{Note that the specific choice quoted in the text ($\hat{ m}_0=10^{-1.4}\eV$) corresponds to the minimum $\hat m_0$ for which we have found no point in the parameter space of the model that would yield large-enough $\eta_B$.} $10^{-1.4}\eV$ see Fig.~\ref{fig:maximum baryon asymmetry 2}. 

Three distinct domains (blue patches in Figs.~\ref{fig:maximum baryon asymmetry 1} and ~\ref{fig:maximum baryon asymmetry 2}) corresponding to different shapes of the RHN spectra can be readily identified :
\begin{figure}[t!]	\centering
	\mbox{
\includegraphics[height=8.5cm,width=8cm]{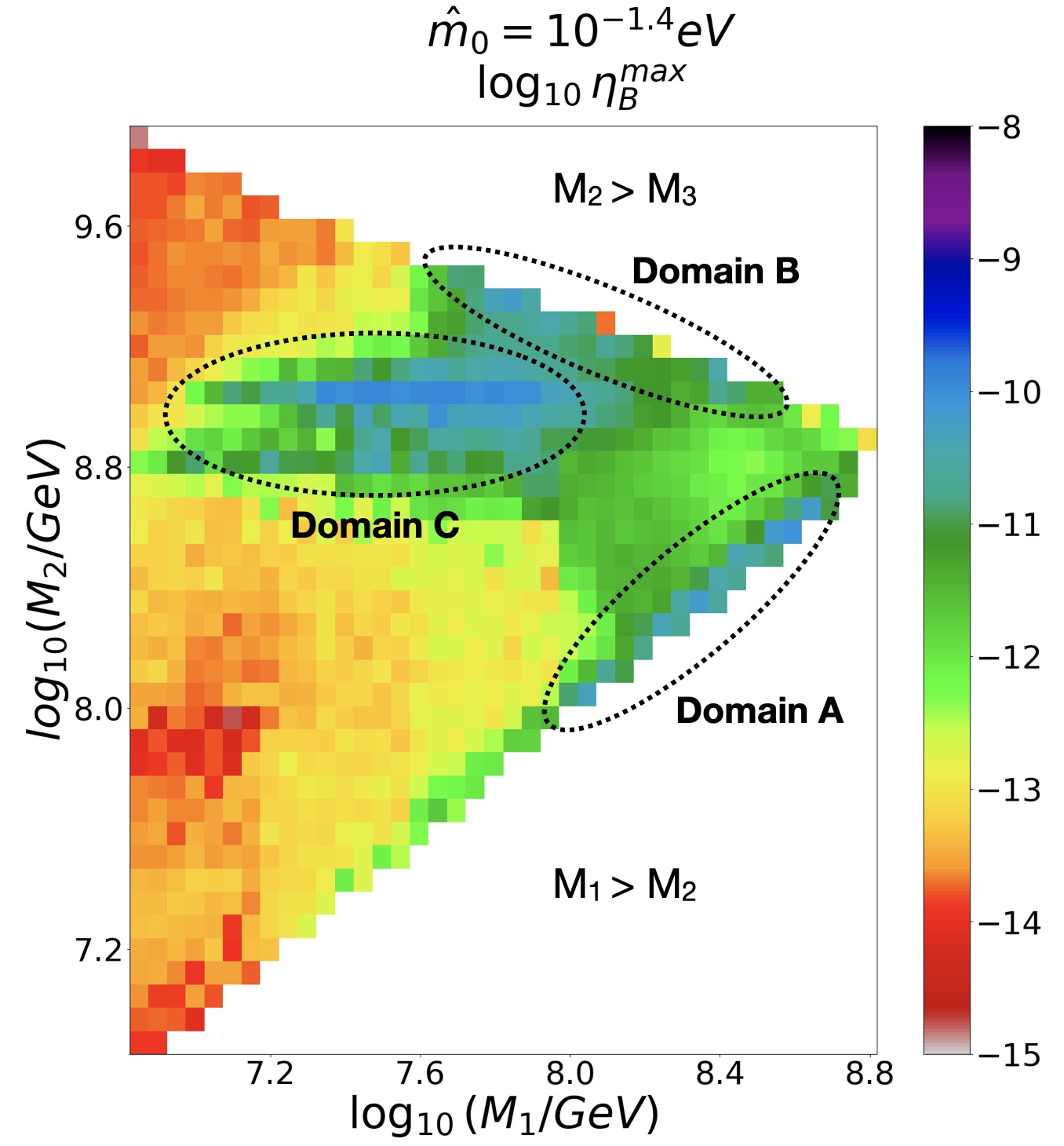}
    }
	\caption{The same as in Fig.~\ref{fig:maximum baryon asymmetry 1} but for $\hat m_0=10^{-1.4}\eV$. 
 Here one can rather clearly distinguish among the three different large-$\eta_B$ domains corresponding to three physically inequivalent regimes, see Section~\ref{sect:Results}. Note, however, that the situation depicted here is not entirely physical as none of the points in the picture yields $\eta_B$ at the level of the observed value of $6\times 10^{-10}$ (the absolute $\eta_B^{\rm max}$ here is a factor of few smaller than that).
 In this sense, the model is incompatible with quasi-degenerate light neutrino spectrum with $\hat m_0\gtrsim 10^{-1.4}\eV$, see also Section~\ref{sec:beta-decay}.
 About 28 milions of points have been generated in preparation of this plot.
	\label{fig:maximum baryon asymmetry 2}
}
\end{figure}
\begin{enumerate}
    \item {\em Domain A} corresponding to the lower-right edge of Fig.~\ref{fig:maximum baryon asymmetry 2} with $M_1$ comparable to $M_2$;
    \item {\em Domain B} corresponding to the upper-right edge of the same plot with $M_2$ comparable to $M_3$ and 
    \item {\em Domain C} ``inside the triangle" of Fig.~\ref{fig:maximum baryon asymmetry 2} characterized by $M_2$ around $10^9\GeV$.
\end{enumerate}
The emergence of each of these regions can be, to a large degree, understood analytically by inspecting the corresponding CP asymmetries and washout factors:
\begin{enumerate}
    \item In {\em Domain A} the resulting $\eta_B$ is dominated by the asymmetry generated in $N_1$ decays and, despite a significant washout, it can still be large enough due to the enhanced $\epsilon_{CP}^1$ of Eq.~\eqref{epsilon i from Y}, see Fig.~\ref{fig:Domain_A_kepsilon}. This case can thus be viewed as a ``standard leptogenesis regime" (albeit out of the Davidson-Ibarra mode and, thus, not entirely trivial to achieve for ``large" $\hat{m}_0$ as in Figs.~\ref{fig:maximum baryon asymmetry 1} and ~\ref{fig:maximum baryon asymmetry 2}) as it occurs close to the edge of the parameter space corresponding to the apex of the triangle defined by Eq.~\eqref{eq:product_rule}. Remarkably, this is  the only domain in which $\eta_B$ can be $N_1$-dominated.
    Note also that for $\hat{m}_0$ smaller than $10^{-4.6}$~eV, this domain grows and $M_1\approx M_2$ relaxes, thereby entering a genuine Davidson-Ibarra mode, see discussion around Equation~\eqref{eq:limit}.
    \item 
    The situation in {\em Domain B} is to a certain degree analogous to that in {\rm Domain A} but for $N_2$-generated asymmetry, which is protected from the subsequent $N_1$-driven washout by the suppression of decoherence effects. Note that for $M_2\sim 10^8\GeV$ one is somewhat below the traditional band of the ``transition regime" of Ref.~\cite{Blanchet:2011xq}. However, the decoherence efficiency is still not large enough to wipe out the entire $\varepsilon_{CP}^2$.
   \item Although leptogenesis in {\em Domain C} is again dominated by $N_2$ decays, the physical picture emerging here is very different from the preceding case -- it relies on a significant washout factor suppression in a region where the CP asymmetry is not particularly enhanced (and, again, decoherence is kept under control). As one can see in the right panel of Fig.~\ref{fig:Domains_BC_kepsilon}, $k_2$ exhibits a very special behavior for $M_2$ around $10^{9}\GeV$ where it suddenly drops to as low as $10^3$ (to be compared to at least $10^5$ outside this domain). In Appendix~\ref{app:minimal_washout} we show that this behavior can be attributed to the analytic structure of the washout factor formula~\eqref{washout_ki} in which the $m_t^2$-proportional contribution can be fully screened for 
\begin{equation}
\label{eq:stripM2}
	M_2\approx \frac{\hat m_{c}^{2}}{\hat m_{0}}\,, 
\end{equation}    
which, indeed, perfectly fits the observation (see also Appendix~\ref{app:inputs}). Note that the slightly chaotic color pattern inside the ``$k_2$ suppression band" in Fig.~\ref{fig:Domains_BC_kepsilon} is an artifact of the steepness of its slopes and a limited computing power stretched over the multidimensional parameter space of the model.
\end{enumerate} 
It is perhaps worth noting that for $\hat{m}_0$ well below the $10^{-1.4}$~eV level that we have predominantly focused on in the discussion above, a band of accidentally small washout defined by Eq.~\eqref{eq:stripM2} would get generally even wider (cf. Fig.~\ref{fig:maximum baryon asymmetry 1}) and, remarkably, a similar band of accidentally small washout would emerge also for $M_1$. 
Moreover, for really small $\hat{m}_0$ (i.e., for $\hat{m}_0\lesssim 10^{-4.6}$~eV, cf. Eq.~\eqref{eq:limit}) yet another set of low washout bands appears for both $M_1$ and $M_2$ at around $\hat{m}_u^2/\hat{m}_0$, see also Section~\ref{sect:parameter_space}.
This, by the way, is also the regime where perturbativity limits discussed in Appendix~\ref{app:perturbativity_further} may become relevant.

Let us conclude this section by reiterating that leptogenesis-compatible parameter-space points have been found only for $\hat m_0\lesssim 10^{-1.4}\eV$ (cf. Figs.~\ref{fig:maximum baryon asymmetry 1} and~\ref{fig:maximum baryon asymmetry 2}) which, in terms of low-scale observables, corresponds to 
\begin{equation}\label{eq:m0limit}
    m_0\lesssim 10^{-1.5}\eV\,.
\end{equation}
This is a robust prediction of the model under consideration and one of the central results of this study. 

\subsection{Neutrino mass limits from beta-decay\label{sec:beta-decay}}
For the effective neutrino mass as measured in beta-decay experiments, the bound~\eqref{eq:m0limit} yields a firm theoretical upper limit on the relevant weighted sum of the neutrino masses, namely,
\begin{equation}
    m_\beta\equiv \sqrt{\sum_{i=1}^3|U_{ei}|^2 m_i^2}\lesssim 0.03\eV,
\end{equation}
which, remarkably enough, is almost a factor of 10 below the design sensitivity of KATRIN, see~\cite{KATRIN:2001ttj,KATRIN:2022ayy}. Hence, should KATRIN see a signal, the model under consideration may be, in principle, refutable in the near future.   
\begin{figure*}
	\centering
	\mbox{
            \includegraphics[height=7.7cm,width=7.5cm]{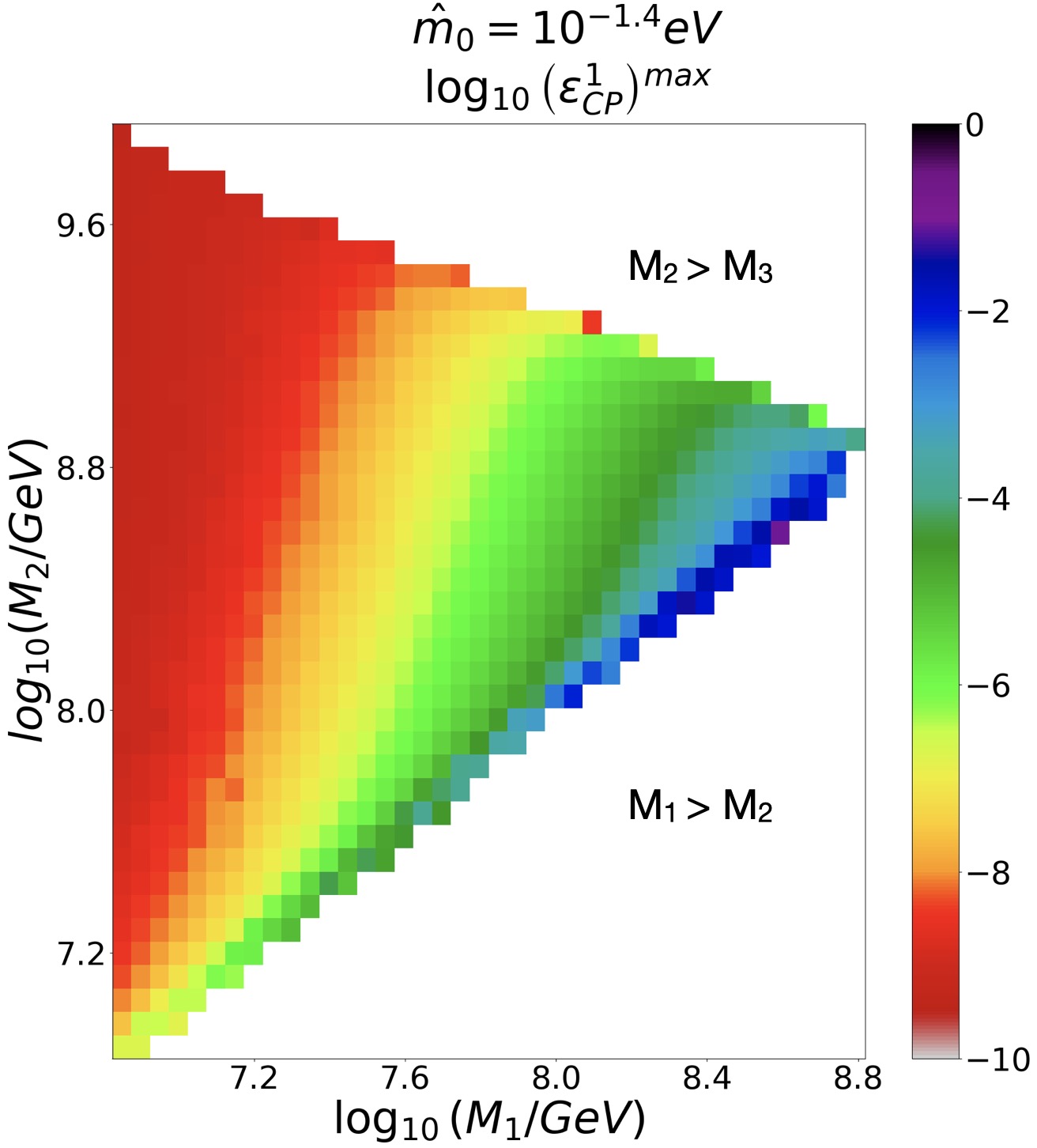}
            \hskip 1cm
            \includegraphics[height=7.7cm,width=7.0cm]{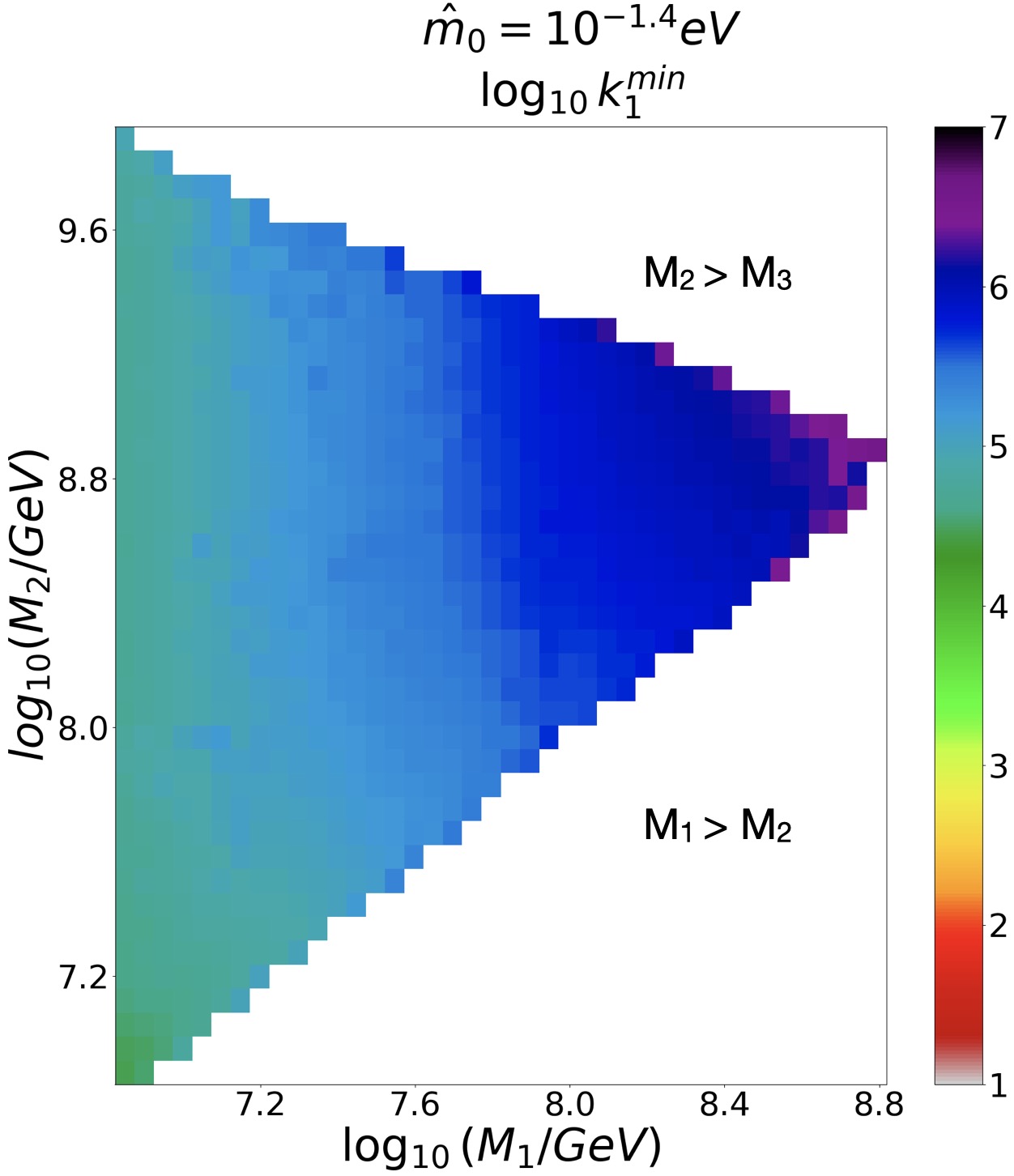}   
        }
    \caption{
The CP asymmetry (left panel) and the washout factor (right panel) driving the $N_1$-dominated $\eta_B$ generation in {\em Domain A} of Fig.~\ref{fig:maximum baryon asymmetry 2}. Note that the washout is never suppressed, so a successful leptogenesis relies on the significant enhancement of $\epsilon_{CP}^1$ in this mode. A vigilant reader may notice that the boundaries of the displayed domains are not ideally straight (which should be the case with log-log plots as implied by the product rule \eqref{eq:product_rule}); this, however, is an artifact of the numerical scanning procedure which, for better performance, was more attentive to certain domains than others; for more details see Appendix~\ref{app:generator}.   
\label{fig:Domain_A_kepsilon}}
\end{figure*}
\begin{figure*}
	\centering
	\mbox{            \includegraphics[height=7.7cm,width=7.5cm]{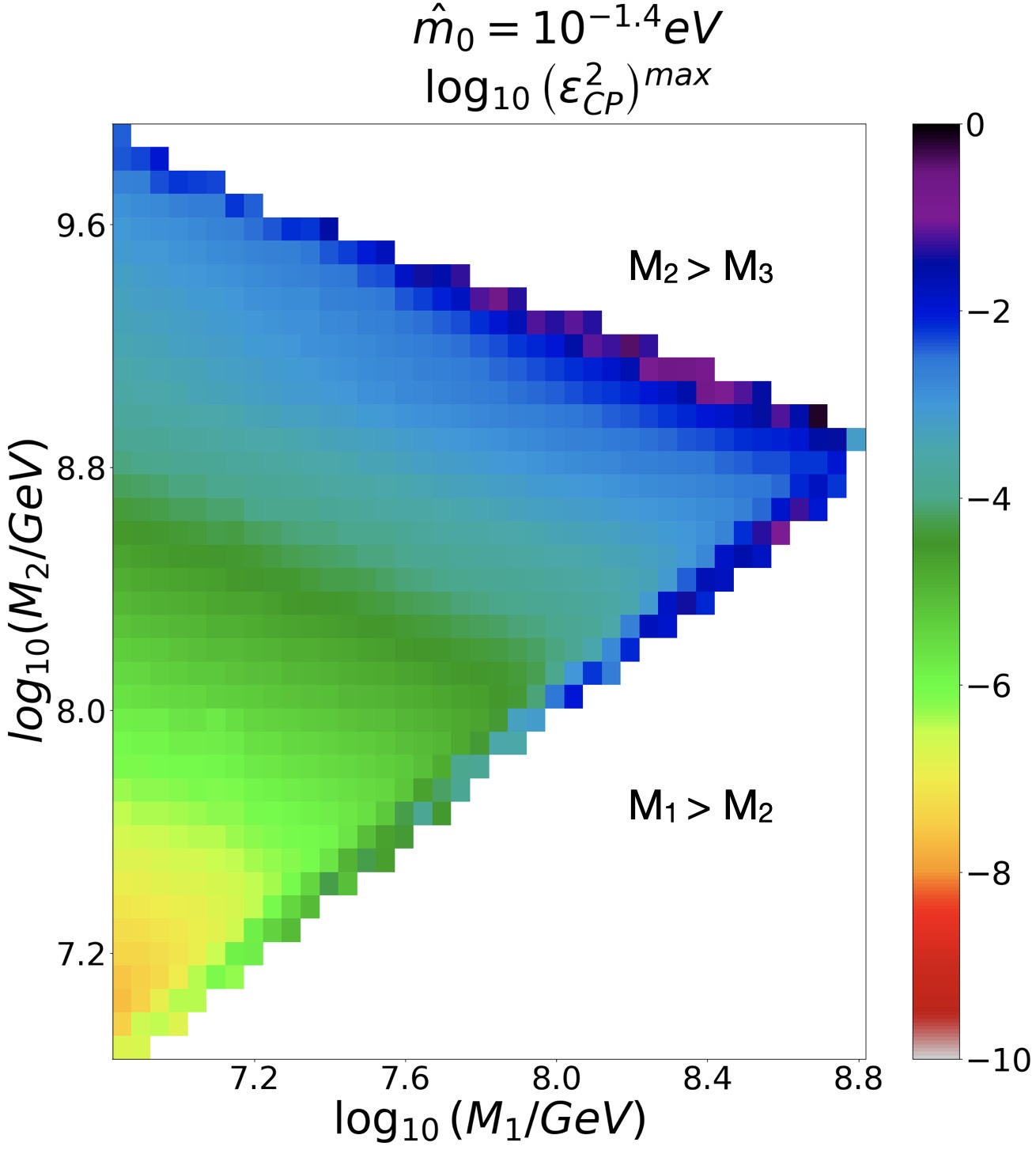}\hskip 1cm 
        \includegraphics[height=7.7cm,width=7.0cm]{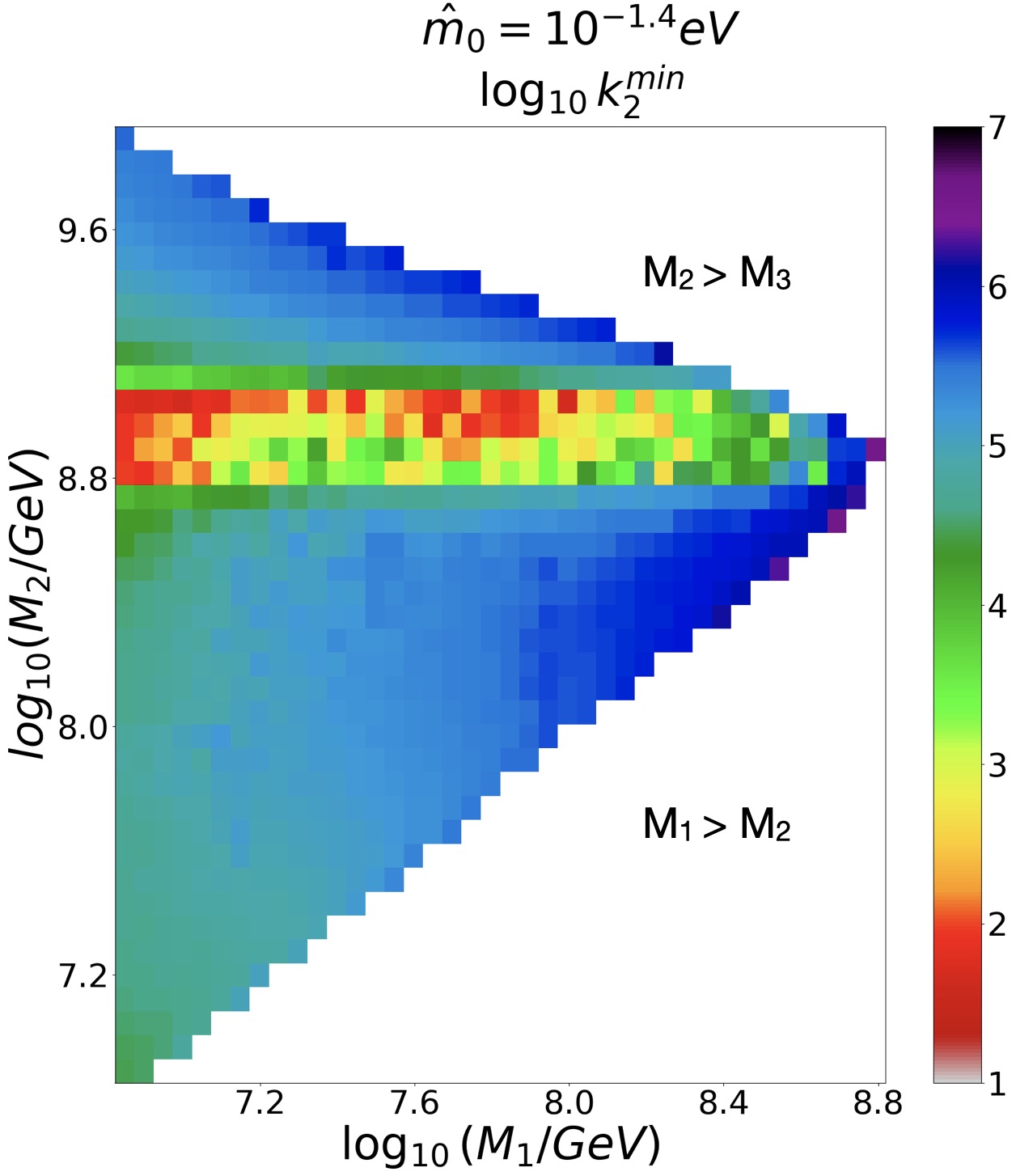}   
        }
    \caption{
The CP asymmetry (left panel) and the washout factor (right panel) driving the $N_2$-dominated $\eta_B$ generation in {\em Domains B and C} of Fig.~\ref{fig:maximum baryon asymmetry 2}. Note that {\em Domain B} corresponds to a large washout compensated by enhanced $|\epsilon_{CP}^2|$ while {\em Domain~C} with moderate $|\epsilon_{CP}^2|$ benefits from a strong washout suppression, see Sect.~\ref{sect:Results} and Appendix~\ref{app:minimal_washout} for the discussion of this feature. In either case, the survival of the $N_2$-generated asymmetries relies heavily on the significant suppression of the decoherence effects. \label{fig:Domains_BC_kepsilon}}
\end{figure*}

\subsection{Leptogenesis-compatible proton decay\label{sect:p-decay}}
Proton decay in the current context has been studied already in Ref.~\cite{Rodriguez:2013rma} (albeit in the simplified case of a real $U_\nu=V_{PMNS}^\dagger U_\ell^L$). Interesting features (such as a lower limit on ${\rm BR}(p^+\to \pi^0 e^+)+{\rm BR}(p^+\to \pi^0 \mu^+)$) have been observed for small $m_0$ while no such limit has been available for $m_0$ above about $10^{-1.5}\eV$, cf. Fig.~5 in Ref.~\cite{Rodriguez:2013rma}.

In the current study the complete CP structure of the model has been invoked, thus providing full generalization of the previous results.
Remarkably, even in such a case the extra constraints from leptogenesis complement those from perturbativity etc. in a way which leads to a pattern similar to that of Ref.~\cite{Rodriguez:2013rma}. Specifically, there is still a lower limit on ${\rm BR}(p^+\to \pi^0 e^+)+{\rm BR}(p^+\to \pi^0 \mu^+)$ for small-enough $m_{0}$ stemming mainly from the lower limit on $M_1$ (as seen in Fig.~\ref{fig:maximum baryon asymmetry 1}), see also Appendix~\ref{app:generator}.  Moreover, the upper limit on $m_{0}$ discussed in the previous Section cuts out exactly those parts of the parameter space where no features in $p$-decay have been previously observed (see again Fig.~5 in Ref.~\cite{Rodriguez:2013rma}). 

\begin{figure}[th!]
    \centering
    \includegraphics[width=8.5cm,height=8.5cm]{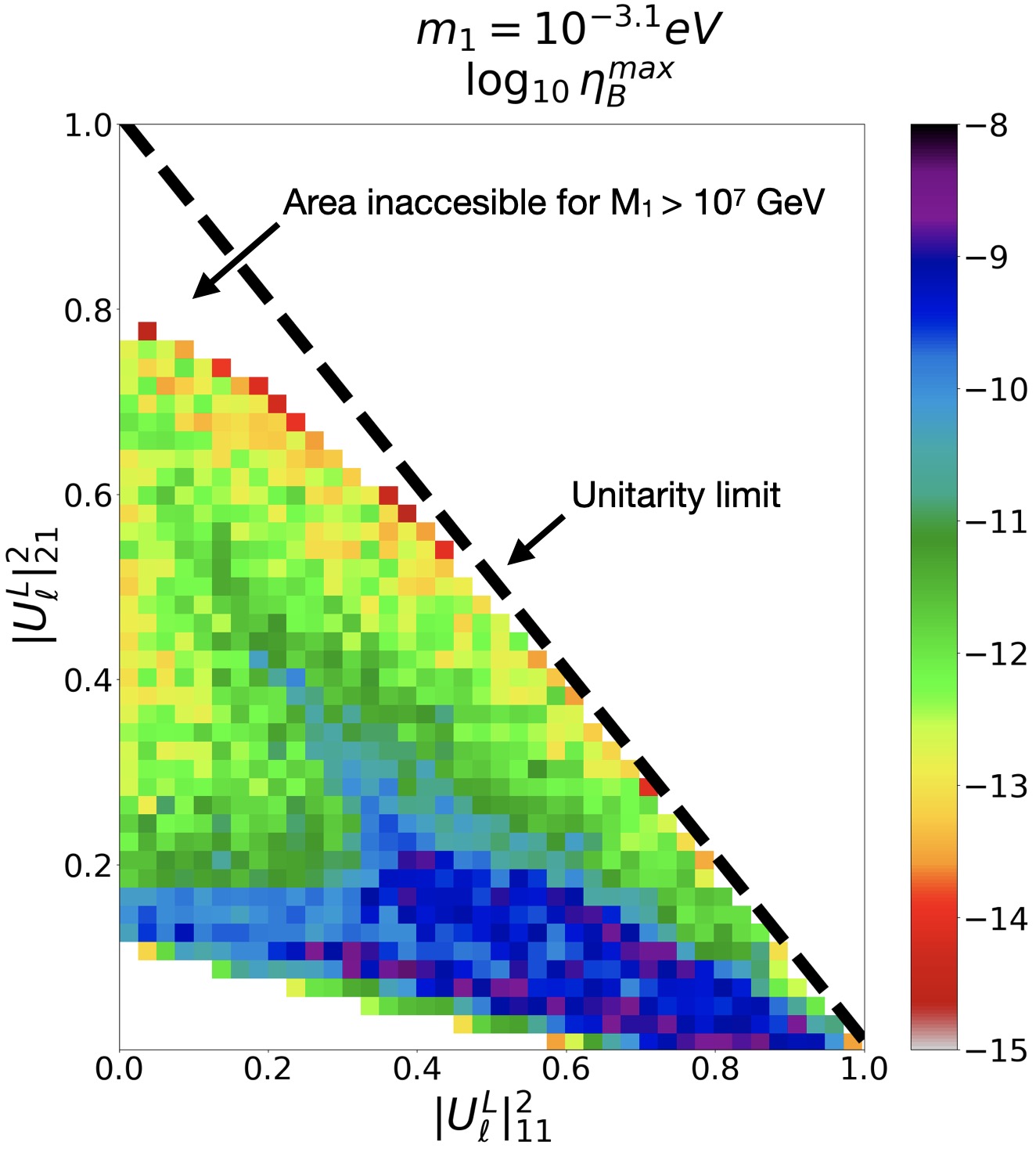}
    \caption{An illustration of the additional constraints leptogenesis can impose on the parameters governing $p$ decay in the current model. In the plot the maxima of $\eta_B$ for different sizes of the two key matrix elements (evaluated at the seesaw scale) driving the $p^+\to \pi^0 e^+$ and $p^+\to \pi^0 \mu^+$ branching ratios are depicted, see e.g. Eq.~\eqref{eq:pdecayformulae}. Two main features are visible here: i) The sum of the two matrix elements can not be arbitrarily suppressed; this is similar to the behaviour seen in the CP conserving case in Ref.~\cite{Rodriguez:2013rma}. ii) There is an upper limit on $|(U_\ell^L)_{21}|^2$ which translates into a previously inaccessible strong upper bound on the individual ${\rm BR}(p^+\to \pi^0 \mu^+)$, see Eq.~\eqref{eq:BRmu}. The white regions within the unitarity domain (below the dashed line) are cut out as they would correspond to $M_1$ below the phenomenological $10^7$~GeV limit, see Section~\ref{sect:parameter_space} and Appendix~\ref{sect:UnufromlimitonM1}, Eq.~\eqref{Ul21_limit}. \label{fig:p-decay}}
\end{figure}

On top of that, successful leptogenesis requires $M_1$ to be larger than about $10^{7.5}$ GeV (cf. Figs.~\ref{fig:maximum baryon asymmetry 1} and~\ref{fig:maximum baryon asymmetry 2}) which translates into a strong constraint on the size of the 21 element of the $U^L_\ell=V_{PMNS} U_\nu$ matrix stemming from inequality~\eqref{leptogenesis 11 condition} of Appendix~\ref{app:generator}
\begin{equation}\label{eq:p_relevant_combination}
\left|\left(U_\nu^*\right)_{k1}^2 m_k\right|\le 2.10^{-4}\left(\frac{10^7\GeV}{M_1}\right){\rm eV}\,,
\end{equation}
as illustrated in Fig.~\ref{fig:p-decay}.
The universal $|(U^L_\ell)_{21}|^2\lesssim 0.8$ bound seen there (common to all $m_0\lesssim 10^{-1.5}\eV$, see also Appendix~\ref{sect:UnufromlimitonM1})
then translates into a limit on the partial 2-body proton decay width into muons 
\begin{equation}
\label{eq:pdecayformulae}
\Gamma(p\to \pi^0 \mu^+)=\frac{1}{2}|(V_{CKM})_{11}|^2|(U^L_\ell)_{21}|^2 \Gamma(p\to \pi^+\overline{\nu}) 
\end{equation} 
in the form of
a global bound on the relevant branching ratio (cf. Eqs.~(5)-(12) in Ref.~\cite{Rodriguez:2013rma}) 
\begin{equation}
\label{eq:BRmuglobal}
   {\rm BR}(p\to \pi^0 \mu^+)\lesssim 0.29\,.
\end{equation}
However, large-enough $\eta_B$ is in practice attained only for a subset of the points conforming the basic condition~\eqref{eq:p_relevant_combination}; for instance, for $m_0\sim 10^{-3}$~eV one has $|(U^L_\ell)_{21}|^2 \lesssim 0.2$ (see again Fig.~\ref{fig:p-decay}) which yields 
\begin{equation}
\label{eq:BRmu}
   {\rm BR}(p\to \pi^0 \mu^+)\lesssim 0.09\,.
\end{equation}
Hence, the strong suppression of proton decay into muons in the ``large" $m_0$ domain can be seen as another potentially testable feature of the current scenario. 

\subsection{Remarks on inverted neutrino hierarchy\label{sect:inversehierarchy}}
The main result of this study, namely, the absence of good points for $m_0$ above about $10^{-1.5}\eV$, see Eq.~\eqref{eq:m0limit}, is a  universal feature shared by both neutrino spectrum types because it is saturated in the quasi-degenerate regime where their difference is not very pronounced.

Concerning the $p$-decay features of the previous subsection, it is again formula~\eqref{eq:p_relevant_combination} that governs the relevant quantities even in the inverted hierarchy case (where $m_0$ corresponds to $m_3$). The only difference is that it connects the $11$ and $21$ elements of the $U_\nu$ matrix rather than its $21$ and $31$ entries as it was the case for the normal hierarchy. Interestingly, the structure of the first row of $V_{PMNS}$ is such that even in this case one still reveals
\begin{equation}
    |(U^L_\ell)_{21}|^2\lesssim 0.8\,,
\end{equation}
albeit for a different reason than for the normal hierarchy, and there is even a similar limit on  $|(U^L_\ell)_{11}|^2$ driving the decay modes into positrons. In any case, the global limit~\eqref{eq:BRmuglobal} remains intact. 
\section{Conclusions}
In this work we have scrutinised the prospects of accommodating thermal leptogenesis within the minimal flipped $SU(5)$ unified model of Ref.~\cite{Rodriguez:2013rma} with the aim to learn more about its flavour structure from the requirement of attaining the measured value of the baryon asymmetry of the Universe. This is by far not a trivial objective as the model prefers a relatively light RH neutrino spectrum (emerging only at two loops) and typically features significant washout effects, at odds with the canonical $N_1$-dominated settings \`a la Davidson and Ibarra.

Nevertheless, as it turns out from a detailed numerical analysis presented in Section~\ref{sect:Results}, thermal leptogenesis does indeed work in this setting, albeit not universally. Among other prerequisites, it generally requires the mass of the lightest active neutrino to be smaller than about $0.03\eV$, well below the target sensitivity limit of KATRIN. In this sense, the model is potentially testable in near future!

Concerning the details of the leptogenesis mechanism operation, several qualitatively different regimes of attaining a realistic $\eta_B$ in the current model have been identified. The first scenario corresponds to the usual $N_1$-dominated asymmetry generation in the situation in which the notorious Davidson-Ibarra limit is evaded by bringing $M_1$ and $M_2$ relatively close to each other (without invoking a resonant behavior for which the model provides no rationale) and the generally large washout factor $k_1$ is at least partly suppressed. In certain situations, however, even the $N_2$-generated asymmetry may be protected from the subsequent $N_1$-driven washout by the suppression of decoherence effects; a large-enough $\eta_B$ can be obtained either due to a strongly enhanced leptonic asymmetry $\epsilon^{2}_{CP}$ or due to an accidentally suppressed $k_2$ washout factor, with both options available within specific parts of the relevant parameter space. 

Remarkably enough, such additional  constraints also significantly improved the limits on the proton decay partial widths studied earlier in Ref.~\cite{Rodriguez:2013rma}. In particular, the strict upper bound on the mass of the lightest active neutrino mass cuts out just those parts of the parameter space where other constraints (such as perturbativity) did not previously provide any valuable insight.

\section*{Acknowledgments}
MM, VM and MZ acknowledge financial support from the Czech Science Foundation (Grantov\'a Agentura \v{C}esk\'e Republiky, GA\v{C}R) via Contract No. 20-17490S, and from the Charles University Research Center Grant No. UNCE/SCI/013. RF acknowledges financial support from the grants PID2019-106087GB-C22 and PID2022-139466NB-C22 funded by MCIN/AEI/10.13039/501100011033 and by “FEDER Una manera de hacer Europa”, as well as from the Junta de Andaluc\'ia grant P21-00199. Special thanks go to Dylan Harries for extensive discussions of various technical aspects of the project in its early phase. 
\appendix

\section{Definitions and conventions\label{app:definitions}}
\subsection{The seesaw anatomy\label{app:seesaw_anatomy}}
In this appendix we fix the notation and conventions used in Sections~\ref{sect:introduction}--\ref{sect:Results}.
The part of the interaction Lagrangian driving the seesaw mechanism reads
\begin{multline}\label{original Lagrangian}
	\mathcal{L}\ni \overline{\ell}_L M_\ell \ell_R+\overline{\nu}_L M_\nu^D \nu_R+\frac{1}{2}\nu_R^T C M_\nu^M \nu_R\\
 -\frac{g}{\sqrt{2}}\overline{\ell}_L\gamma^\mu\nu_L W_\mu^-+\text{H.c.}\,,
\end{multline}
where $\nu_L$, $\nu_R$, $\ell$, and $W_\mu$ are left-handed neutrino, right-handed neutrino, charged lepton, and W-boson fields, respectively, $M_\ell$ and $M_\nu^D$ stand for the Dirac mass matrices of charged leptons and neutrinos, and $M_\nu^M$ denotes the heavy sector Majorana mass matrix. 

As usual, introducing $n_L^T=(\nu_L, \nu_R^c
)^T$, the heavy neutrino part of~\eqref{original Lagrangian} can be readily rewritten in a compact form 
\begin{equation}
\overline{\nu}_L M_\nu^D \nu_R+\frac{1}{2}\nu_R^T C M_\nu^M \nu_R+\text{H.c.}=\frac{1}{2}n_L^T C\mathcal{M}\/ n_L+\text{H.c.}\,,\nonumber
\end{equation}
provided
\begin{equation}
	\mathcal{M}=\begin{pmatrix}
		0 & (M_\nu^D)^*\\
		(M_\nu^D)^\dagger & (M_\nu^M)^*
	\end{pmatrix}.
\end{equation}
For $|M_\nu^M|\gg |M_\nu^D|$, its partial diagonalization yields an effective low-energy lagrangian in the form 
\begin{equation}
	\mathcal{L}\ni \overline{\ell}_L M_\ell \ell_R+\frac{1}{2}\nu^T C m_{LL}^* \nu-\frac{g}{\sqrt{2}}\overline{\ell}_L\gamma^\mu\nu W_\mu^-+\text{H.c.}.
\end{equation}
where
the effective light neutrino ($\nu\approx \nu_L$) mass matrix obeys
\begin{equation}\label{original mLL app}
	m_{LL}^*=-(M_\nu^D)^* ({M_\nu^{M}}^{*})^{-1} (M_\nu^D)^\dagger,
\end{equation}
and all heavy degrees of freedom have been integrated out.
The low-energy sector is then fully diagonalized by unitary transformations
\begin{equation}
	\nu'=U_\nu \nu, \qquad \ell'_L = U_\ell^L \ell_L, \qquad \ell'_R = U_\ell^R \ell_R,
\end{equation}
which yield
\begin{align}
	 M_\ell&=(U_\ell^L)^\dagger D_\ell U_\ell^R\,,\\
	  m_{LL}^*&=U_\nu^T D_\nu U_\nu\,, \label{diagonalization mLL app}\\
	  V_{PMNS}&=U_\ell^L U_\nu^\dagger\,, \label{PMNS app}
\end{align}
where $D_\ell$ and $D_\nu$ denote the real positive diagonal charged lepton and light neutrino mass matrices, respectively, and $V_{PMNS}$ is the Pontecorvo-Maki-Nakagawa-Sakata matrix\footnote{Strictly speaking, the combination~\eqref{PMNS app} defines the PMNS matrix in the `raw' form, i.e., prior to striping the three extractable unphysical LHS phases $P_\ell$ in the general decomposition
\begin{equation}\label{PMNS app2}
   V_{PMNS}=P_\ell V'_{PMNS}P_\nu
\end{equation} where $P_\nu={\rm diag}(e^{i\alpha_1},e^{i\alpha_2},1)$ contains the two physical Majorana phases and $V'_{PMNS}$ stands for the ``stripped" (i.e., CKM-like) form of the same depending on three angles and the remaining Dirac phase.}.
In a similar way, one can diagonalize the mass matrix of the right-handed neutrinos
\begin{equation}\label{diagonalization MnuM}
	M_\nu^M=\tilde{U}^T D_\nu^M \tilde{U}.
\end{equation}
which, by virtue of~(\ref{original mLL app}) and~(\ref{diagonalization mLL app}) can be rewritten in the form
\begin{equation}\label{see-saw app}
	\tilde{U}^T D_\nu^M \tilde{U} = - \hat{D}_u U_\nu^T D_\nu^{-1} U_\nu \hat{D}_u\,. 
\end{equation}
This is the central formula of our interest, cf. Section~\ref{sect:seesaw}.
\subsection{Parameter counting in the $M_\nu^M$-diagonal basis\label{App:Basis change}}
Let us count the independent quantities in terms of which one can fully parametrize all structures entering the $\eta_B$ calculation of Section~\ref{sec:SU5leptogenesis}, namely, $M_\nu^M$ of~\eqref{seesaw_diagonal_form} (see also~\eqref{diagonalization MnuM} and~\eqref{see-saw app}) and $Y_\nu$ of~\eqref{Yukawa_from_U}.

In the basis in which the RHN mass matrix $M_\nu^M$ is diagonal the Yukawa matrix reads 
\begin{equation}
	Y_\nu=\tfrac{1}{v}U_\ell^L M_\nu^D \tilde{U}^\dagger=\tfrac{1}{v}V_{PMNS}U_\nu \hat{D}_u \tilde{U}^\dagger,
\end{equation}
where $\tilde{U}$ is defined by Eq.~\eqref{diagonalization MnuM}.
Plugging~\eqref{see-saw app} therein one obtains
\begin{equation}
\label{eq:first_step}
\tilde{U}^T D_\nu^M \tilde{U}=- \hat{D}_u P_U^R {U'_\nu}^T P_U^L D_\nu^{-1} P_U^L U'_\nu P_U^R \hat{D}_u
\end{equation}
where the $U_\nu$ matrix of~\eqref{diagonalization mLL app} has been decomposed as 
\begin{equation}
    \label{Unudecomposition}
    U_\nu\equiv P_U^L U'_\nu P_U^R
\end{equation}
with $P_U^L$ and $P_U^R$ denoting diagonal matrices of `extractable' phases (two in $P_U^L$ and three in $P_U^R$) and $U'_\nu$ stands for the remainder of this decomposition, i.e. a unitary $3\times 3$ matrix with $3$ angles and one remaining nonextractable phase. 
Using the commutativity of $P_U^R$ and $D_u$ the $P_U^R$ phases can be pushed to the LHS of~\eqref{eq:first_step} 
\begin{equation}
\label{eq:second_step}
{P_U^R}^*\tilde{U}^T D_\nu^M \tilde{U}{P_U^R}^*=- \hat{D}_u {U'_\nu}^T P_U^L D_\nu^{-1} P_U^L U'_\nu \hat{D}_u.
\end{equation}
At the same time, the $Y_\nu$ can be written as 
\begin{equation}
	Y_\nu=\tfrac{1}{v}V_{PMNS} P_U^L U'_\nu \hat{D}_u P_U^R \tilde{U}^\dagger,
\end{equation}
from where one can see that $\tilde{U}$ enters both these relations always together with $P_U^R$. Hence, it is sufficient to work with their compound $\tilde{U}_1\equiv \tilde{U} {P_U^R}^*$ which, however, can be fully determined from the variant of Eq.~\eqref{eq:second_step} in the form 
\begin{equation}
\label{eq:DMU1}
\tilde{U}_1^T D_\nu^M \tilde{U}_1=- \hat{D}_u {U'_\nu}^T P_U^L D_\nu^{-1} P_U^L U'_\nu \hat{D}_u\,.
\end{equation}
Hence, it is a function of only $P_U^L$ and $D_\nu$ (taking $D_u$ as a known constant matrix), i.e. 2 unknown phases of 
\begin{equation}\label{def:phases L}
	P_U^L=\diag\{e^{i\beta_1/2}, e^{i\beta_2/2},1\}
\end{equation}
and 3 unknown angles and 1 phase in $U'_\nu$.
At the same time, $Y_\nu$ is rewritten as
\begin{equation}
    \label{eq:YnuU1}
	Y_\nu=\tfrac{1}{v}V_{PMNS} P_U^L U'_\nu \hat{D}_u \tilde{U}_1^\dagger,
\end{equation}
and, barring the three non-physical phases one can discard from the LHS of the `raw' form of $V_{PMNS}$ matrix of Eq.~\eqref{PMNS app2}, it depends only on $P_U^L$, $U_\nu'$ and the remaining $V_{PMNS}$ parameters (i.e. 3 known mixing angles, the (more or less constrained) Dirac phase and two unknown low-energy Majorana phases of Eq.~\eqref{PMNS app2}).    

Hence, taking into account the mass of the lightest SM neutrinos ($m_0$) in terms of which $D_\nu$ can be fully\footnote{This is true up to the uncertainties in the determination of neutrino mass square differences which, however, do not play any significant role in the analysis of Sect.~\ref{sect:Results}.} reconstructed, one counts the total of 9 free parameters ($m_0$, $\alpha_{1,2}$, $P_U^L$, $U_\nu'$) in terms of which all the leptogenesis-relevant structures can be constructed via Eqs.~\eqref{eq:DMU1} and~\eqref{eq:YnuU1}. 

\section{Experimental inputs\label{app:inputs}}
For the neutrino sector observables we use the information provided in~\cite{deSalas:2020pgw}, namely  
\begin{equation}
	\label{neutrino experiment}
	\begin{split}
		\Delta m_{21}^2&=7.50\times 10^{-5}\eV^2, \
	|\Delta m_{31}^2|=2.55\times 10^{-3}\eV^2\, \text{(NH)},\\
	\Delta m_{21}^2&=7.50\times 10^{-5}\eV^2, \
	|\Delta m_{31}^2|=2.45\times 10^{-3}\eV^2\, \text{(IH)},
	\end{split}
\end{equation}	
Concerning the PMNS matrix elements \eqref{PMNS}, we neglect the small uncertainties in the mixing angles and, for the sake of simplicity (and without much of an impact onto the results, see Section~\ref{sec:SU5leptogenesis} and Appendix~\ref{app:generator_phases}), fix the Dirac CP phase at its central value. For the normal hierarchy, this means
\begin{equation}\label{VPMNS}
\theta_{12}\approx 34.3^{\circ}, \
\theta_{23}\approx 49.26^{\circ}, \
\theta_{13}\approx 8.53^{\circ}, \
\delta \approx 1.08\, \pi,
\end{equation}
whereas for the inverted case, one has
\begin{equation} \label{VPMNS inverted}
	\theta_{12}\approx 34.3^{\circ}, \
        \theta_{23}\approx 49.46^{\circ}, \
	\theta_{13}\approx 8.58^{\circ}, \
	\delta \approx 1.58\, \pi.
\end{equation}
However, these data (especially the masses) must be evolved to the $M_5$ scale before getting plugged into formulae like~\eqref{seesaw_diagonal_form} (and/or to the seesaw scale for leptogenesis-related calculations).
According to \cite{Antusch2003}, the mass effects may be to a good degree approximated by universal multiplicative factors, e.g.,
\begin{equation}\label{neutrino running}
    r=\frac{m_\nu(M_{BL})}{m_\nu (1\,{\rm eV})}\equiv\frac{\hat{m}_\nu}{m_\nu (1\,{\rm eV})} \approx 1.32.
\end{equation}
where the hat denotes seesaw-scale quantities, see Section~\ref{sect:seesaw}.

For the up-type quark masses, the second important ingredient of the central formula~\eqref{seesaw_diagonal_form}, we use the results of~\cite{Xing2008}, namely    
\begin{equation}
\label{eq:upquarkratios}
	\frac{m_t(M_5)}{m_t(M_Z)}\approx 0.44, \quad
	\frac{m_c(M_5)}{m_c(M_Z)}\approx 0.38, \quad
	\frac{m_u(M_5)}{m_u(M_Z)}\approx 0.38.
\end{equation}
with 
\begin{align}
\label{eq:upquarkMZ}
	m_t(M_Z) &\doteq 173 {\,\rm GeV}, \;
	m_c(M_Z)\doteq 0.62 {\,\rm GeV}, \nonumber\\
	 m_u(M_Z)& \doteq 1.3 {\,\rm MeV}.
\end{align}
Note, however, that in formulae like~\eqref{eq:product_rule} one in fact needs the unification-scale masses of Eqs.~\eqref{eq:upquarkratios} and~\eqref{eq:upquarkMZ} to be evolved down to the seesaw scale by means of RGEs for the neutrino Dirac mass matrix; for these parameters (denoted by hats in Eq.~\eqref{eq:product_rule}) we take
\begin{equation}
\label{eq:upquarkMBL}
	\hat{m}_t \doteq 72 {\,\rm GeV}, \;\;
	\hat{m}_c\doteq 0.23 {\,\rm GeV},\;\;
	 \hat{m}_u \doteq 0.48 {\,\rm MeV}.
\end{equation}
With all this at hand, one can finally evaluate the RHS of Equation \eqref{eq:product_rule}
\begin{equation}\label{constraint product}
	\hat{m}_1 \hat{m}_2 \hat{m}_3 M_1 M_2 M_3=\hat{m}_u^2 \hat{m}_c^2 \hat{m}_t^2\approx 6.3\times 10^{49}\eV^6.
\end{equation}


\section{Scalar potential convexity constraints\label{app:global_features}}

In this Appendix we shall address a natural question whether the scalar spectrum of the model and/or the other scalar parameters entering the RH neutrino mass formula~\eqref{eq:neutrino formula} may be further constrained from the very natural requirement of the scalar potential stability. By this we mean its boundedness from below in all scalar field directions which is necessary to prevent a runaway behaviour of certain scalar modes.  

Remarkably enough, the answer to this question is negative. In technical terms, it may be shown that for every shape of the triplet mass spectrum $m_{\Delta_i}$ and for every $U_\Delta$ matrix diagonalizing the triplet masses there always exists a point in the parameter space of the model for which the scalar potential is bounded from below as required.
\subsection{Higgs-doublet diagonal basis\label{app:pentuplet_basis}}
The statement above is best proven in a special basis corresponding to a rotation in the two scalar pentuplet field space 
\begin{equation}\label{eq:rotation}
\Phi_{5,i}\rightarrow U_{ij}\Phi_{5,j}\,,
\end{equation}
which renders the Higgs doublet mass matrix diagonal {\em by definition}. As we shall see, this is also a very convenient basis for  the analysis of  perturbativity constraints imposed on $M^M_\nu$ in Appendix~\ref{app:perturbativity}.    

Let us start by recalling the scalar potential of the two pentuplet model as written in Appendix~B.2 of Ref.~\cite{Harries:2018tld}, namely,
\begin{equation}
V=V_{2}+V_{3}+V_{4}\,,
\end{equation}
with
\begin{align}
V_{2} & =\tfrac{1}{2}m_{10}^{2}\textrm{Tr}\left(\Phi_{10}^{\dagger}\Phi_{10}\right)+m_{5}^{2}\Phi_{5}^{\dagger}\Phi_{5}+m_{5'}^{2}\Phi_{5}^{\prime\dagger}\Phi_{5}^{\prime}\nonumber \\
\label{eq:V2}
 & +\left(m_{12}^{2}\Phi_{5}^{\dagger}\Phi_{5}^{\prime}+\textrm{H.c.}\right)\,,\\
V_{3} & =\frac{\mu}{8}\epsilon^{ijklm}\left(\Phi_{10}\right)_{ij}\left(\Phi_{10}\right)_{kl}\left(\Phi_{5}\right)_{m} \nonumber\\
 & +\frac{\mu'}{8}\epsilon^{ijklm}\left(\Phi_{10}\right)_{ij} \left(\Phi_{10}\right)_{kl}\left(\Phi_{5}^{\prime}\right)_{m}+\textrm{H.c.}\,,\\
V_{4} & =\frac{1}{4}\lambda_{1}\left[\textrm{Tr}\left(\Phi_{10}^{\dagger}\Phi_{10}\right)\right]^{2}+\frac{1}{4}\lambda_{2}\textrm{Tr}\left(\Phi_{10}^{\dagger}\Phi_{10}\Phi_{10}^{\dagger}\Phi_{10}\right)\nonumber \\
 & +\lambda_{3}\left(\Phi_{5}^{\dagger}\Phi_{5}\right)^{2}+\tilde{\lambda}_{3}\left(\Phi_{5}^{\prime\dagger}\Phi_{5}^{\prime}\right)^{2}+\lambda_{6}\left(\Phi_{5}^{\dagger}\Phi_{5}^{\prime}\right)\left(\Phi_{5}^{\prime\dagger}\Phi_{5}\right)\nonumber \\
 & +\tilde{\lambda}_{6}\left(\Phi_{5}^{\dagger}\Phi_{5}\right)\left(\Phi_{5}^{\prime\dagger}\Phi_{5}^{\prime}\right)+\frac{1}{2}\lambda_{4}\Phi_{5}^{\dagger}\Phi_{5}\textrm{Tr}\left(\Phi_{10}^{\dagger}\Phi_{10}\right)\nonumber \\
 & +\frac{1}{2}\tilde{\lambda}_{4}\Phi_{5}^{\prime\dagger}\Phi_{5}^{\prime}\textrm{Tr}\left(\Phi_{10}^{\dagger}\Phi_{10}\right)+\lambda_{5}\Phi_{5}^{\dagger}\Phi_{10}\Phi_{10}^{\dagger}\Phi_{5}\nonumber \\
 & +\tilde{\lambda}_{5}\Phi_{5}^{\prime\dagger}\Phi_{10}\Phi_{10}^{\dagger}\Phi_{5}^{\prime}+\Big[\eta_{1}\left(\Phi_{5}^{\dagger}\Phi_{5}\right)\left(\Phi_{5}^{\dagger}\Phi_{5}^{\prime}\right)\nonumber \\
 & +\eta_{2}\left(\Phi_{5}^{\dagger}\Phi_{5}^{\prime}\right)^{2}+\eta_{3}\left(\Phi_{5}^{\dagger}\Phi_{5}^{\prime}\right)\left(\Phi_{5}^{\prime\dagger}\Phi_{5}^{\prime}\right)+\frac{1}{2}\lambda_{7}\Phi_{5}^{\dagger}\Phi_{5}^{\prime}\nonumber \\
 & \times\textrm{Tr}\left(\Phi_{10}^{\dagger}\Phi_{10}\right)+\lambda_{8}\Phi_{5}^{\dagger}\Phi_{10}\Phi_{10}^{\dagger}\Phi_{5}^{\prime}+\textrm{H.c.}\Big]\,,\label{eq:V4}
\end{align}
where the $\tfrac{1}{2}$ factor in front of $m_{10}^2$ in Eq.~\eqref{eq:V2}, shared also by the relevant kinetic term
\begin{equation} \label{eq:kinetic}
\mathcal{L}_{\text{kin.}} = \frac{1}{2} (D^\mu
  \Phi_{10})^\dagger_{ij} (D_\mu \Phi_{10})^{ji}
  + (D^\mu \Phi_5)^{\dagger}_{i} (D_\mu \Phi_5)^{i}\,,
\end{equation}
accounts for the symmetries of the defining fields
\begin{equation} \label{eq:scalar-field-components}
  \Phi_{10} = \begin{pmatrix}
    0 & \overline{D}_3 & -\overline{D}_2 & U^1 & D^1 \\
    -\overline{D}_3 & 0 & \overline{D}_1 & U^2 & D^2 \\
    \overline{D}_2 & -\overline{D}_1 & 0 & U^3 & D^3 \\
    -U^1 & -U^2 & -U^3 & 0 & \Sigma \\
    -D^1 & -D^2 & -D^3 & -\Sigma & 0 \\
  \end{pmatrix} , \;
  \Phi_5\!\mbox{}^{(}\mbox{}'\mbox{}^{)} = \begin{pmatrix}
    T^1 \\
    T^2 \\
    T^3 \\
    H^0 \\
    H^-
  \end{pmatrix}\mbox{}\!\mbox{}^{(}\mbox{}'\mbox{}^{)}
\end{equation}
and ensures that $m_{10}^2$ corresponds to the tree-level masses (squared) of the correctly normalized components of $\Phi_{10}$ in the unbroken phase. 

The SM $SU(2)_L$-doublet and $SU(3)_c$-triplet mass matrices in the broken phase (devised by $\langle\Sigma\rangle\equiv V_{G}$
responsible for breaking the $SU(5)\times U(1)_{X}$ symmetry down to the $SU(3)_{C}\times SU(2)_{L}\times U(1)_{Y}$ of the SM) then read
\begin{equation}
\label{eq:doublet_matrix}
M_{H}^{2}=\begin{pmatrix}m_{5}^{2}+V_{G}^{2}\left(\lambda_{4}+\lambda_{5}\right) & m_{12}^{2}+V_{G}^{2}\left(\lambda_{7}+\lambda_{8}\right)\\
m_{12}^{2*}+V_{G}^{2}\left(\lambda_{7}^{*}+\lambda_{8}^{*}\right) & m_{5^{\prime}}^{2}+V_{G}^{2}\left(\tilde{\lambda}_{4}+\tilde{\lambda}_{5}\right)
\end{pmatrix}
\end{equation}
in the $\left(H,H^{\prime}\right)$ basis and
\begin{equation}\label{eq:triplet_matrix}
M_{\Delta}^{2}=\begin{pmatrix}-\lambda_{2}V_{G}^{2} & \mu V_{G} & \mu^{\prime}V_{G}\\
\mu^{*}V_{G} & m_{5}^{2}+\lambda_{4}V_{G}^{2} & m_{12}^{2}+\lambda_{7}V_{G}^{2}\\
\mu^{\prime*}V_{G} & m_{12}^{2*}+\lambda_{7}^{*}V_{G}^{2} & m_{5^{\prime}}^{2}+\tilde{\lambda}_{4}V_{G}^{2}
\end{pmatrix}
\end{equation}
in the $\left(\overline{D}^{\dagger},T,T^{\prime}\right)$ basis, respectively.
The freedom to rotate the two 5-plets as in Eq.~\eqref{eq:rotation}
can be used to diagonalize the $M_{H}^{2}$ matrix~\eqref{eq:doublet_matrix}, i.e., zero-out its off-diagonal element, which means\footnote{For the sake of simplicity here we do not make any notational distinction between the values of the same parameters in the two different bases.}
\begin{align}
m_{12}^{2}+V_{G}^{2}\left(\lambda_{7}+\lambda_{8}\right) & =0\,, \label{eq:pent_diag_1}\\
m_{5}^{2}+V_{G}^{2}\left(\lambda_{4}+\lambda_{5}\right) & \equiv m_{h}^{2}\,,\\
m_{5^{\prime}}^{2}+V_{G}^{2}\left(\tilde{\lambda}_{4}+\tilde{\lambda}_{5}\right) & \equiv m_{H}^{2} \label{eq:pent_diag_2}\,.
\end{align}
Note that by adjusting $m_{12}^{2}$, $m_{5}^{2}$ and $m_{5^{\prime}}^{2}$ the
above relations can be met for any masses $m_{h}^{2}$ and $m_{H}^{2}$,
as well as for any values of the quartic couplings. 
Note that in this basis the SM Higgs doublet couples only to one of the $\{Y_{10}$, $ Y'_{10}\}$ matrices; this observation will indeed become handy in Appendix~\ref{app:perturbativity}.

Note also that in the same basis the triplet mass matrix~\eqref{eq:triplet_matrix} can be written in a simplified form (i.e. with only two rather than four entries with more than a single summand) as
\begin{equation}\label{eq:triplet_matrix_reduced}
M_{\Delta}^{2}=\left(\begin{array}{ccc}
-\lambda_{2}V_{G}^{2} & \mu V_{G} & \mu^{\prime}V_{G}\\
\mu^{*} V_{G} & m_{h}^{2}-\lambda_{5}V_{G}^{2} & -\lambda_{8}V_{G}^{2}\\
\mu^{\prime*}V_{G} & -\lambda_{8}^{*}V_{G}^{2} & m_{H}^{2}-\widetilde{\lambda}_{5}V_{G}^{2}
\end{array}\right)\,.
\end{equation}
It is hardly surprising that there is still enough freedom for the spectrum of this hermitian matrix to assume any shape, irrespective of the specific choice of 
$m_{h}^{2}$, $m_{H}^{2}$ and $V_{G}$ above - indeed the correlation between the two structures in Eqs.~\eqref{eq:doublet_matrix} and~\eqref{eq:triplet_matrix} is rather loose.
\vskip 1cm


\subsection{Asymptotic convexity of the scalar potential\label{app:convexity}}
However, can the scalar potential be bounded from below (or perhaps even asymptotically convex) for any choice
of the physical parameters $m_{h}^{2}$, $m_{H}^{2}$ and $M_{\Delta}^{2}$ (with the last one encoding both the triplet spectrum $m_{\Delta_i}$ and the corresponding unitary diagonalization matrix $U_\Delta$)?
The answer to this question is affirmative in the sense that for any choice
of the quartic parameters $\lambda_{2}$, $\lambda_{5}$, $\widetilde{\lambda}_{5}$
and $\lambda_{8}$ in Eq.~\eqref{eq:triplet_matrix_reduced} it is always possible to pick the remaining couplings
of $V_{4}$ in Eq.~\eqref{eq:V4} such that $V_{4}>0$ for any non-null
field configuration. In fact, any possible instability of the $V_4$ substructure 
\begin{align}
& \frac{\lambda_{2}}{4}\textrm{Tr}\left(\Phi_{10}^{\dagger}\Phi_{10}\Phi_{10}^{\dagger}\Phi_{10}\right)
+\lambda_{5}\Phi_{5}^{\dagger}\Phi_{10}\Phi_{10}^{\dagger}\Phi_{5}\nonumber \\
&+\widetilde{\lambda}_{5}\Phi_{5}^{\prime\dagger}\Phi_{10}\Phi_{10}^{\dagger}\Phi_{5}^{\prime}+\left(\lambda_{8}\Phi_{5}^{\dagger}\Phi_{10}\Phi_{10}^{\dagger}\Phi_{5}^{\prime}+\textrm{H.c.}\right) \label{eq:V4_suspect_part1}
\end{align}
due to a ``sub-optimal" choice of $\lambda_{2}$, $\lambda_{5}$, $\widetilde{\lambda}_{5}$
and $\lambda_{8}$
can be resolved by adequately
tuning the values of $\lambda_{1}$, $\lambda_{3}$ and $\widetilde{\lambda}_{3}$ of
\begin{equation}
\label{eq:V4_suspect_part2}
    \frac{\lambda_{1}}{4}\left[\textrm{Tr}\left(\Phi_{10}^{\dagger}\Phi_{10}\right)\right]^{2}
    +
    \lambda_{3}\left(\Phi_{5}^{\dagger}\Phi_{5}\right)^{2}
    +
    \widetilde{\lambda}_{3}\left(\Phi_{5}^{\prime\dagger}\Phi_{5}^{\prime}\right)^{2}
\end{equation}
so that the two structures~\eqref{eq:V4_suspect_part1} and~\eqref{eq:V4_suspect_part2} taken together form a positively definite form; for this it is sufficient to fulfill
\begin{align}
\lambda_{1}> & \tfrac{1}{2}\left|\lambda_{2}\right|+2\left|\lambda_{5}\right|+2\left|\widetilde{\lambda}_{5}\right|+4\left|\lambda_{8}\right|\,,\\
\lambda_{3}> & \tfrac{1}{2}\left|\lambda_{5}\right|+\tfrac{1}{2}\left|\lambda_{8}\right|\,,\\
\widetilde{\lambda}_{3}> & \tfrac{1}{2}\left|\widetilde{\lambda}_{5}\right|+\tfrac{1}{2}\left|\lambda_{8}\right|\,.
\end{align}
Hence, the scalar potential can be always made convex (and, thus, bounded from below) regardless of the specific form of the doublet and triplet scalar spectrum and the $U_\Delta$ unitary matrix governing the central formula~\eqref{eq:neutrino formula}.

\section{Perturbativity of RH neutrino masses\label{app:perturbativity}}

In Appendix~\ref{app:convexity} it was demonstrated that even with the extra condition of the scalar potential convexity, the triplet spectrum encoded in $m_{\Delta_i}$ in equation \eqref{eq:neutrino formula} can still be taken as essentially arbitrary along with an arbitrary unitary mixing matrix $U_\Delta$.
At the same time (cf. Appendix~\ref{app:pentuplet_basis}), the basis in the $\Phi_5-\Phi'_5$ plane can be chosen in such a way that one of the two pentuplets fully contains the Standard Model Higgs doublet and there is no projection of the same onto the other one. This means that the first pentuplet (and hence the triplet component within) couples to the matter currents emerging from the $\Yten \TenFermion \TenFermion \FiveHiggs$ term in Eq.~\eqref{eq:two-five-yukawa-sector} through the Yukawa coupling of the SM Higgs, i.e. the down-quark Yukawa matrix $Y_d\equiv 8 Y_{10}$, see Section~III.B.1 in Ref.~\cite{Harries:2018tld}. Hence, the equation \eqref{eq:neutrino formula} can be rewritten as
\begin{equation}\label{eq:neutrino_formula_Yd}
        \begin{split}
		M_\nu^M=\frac{6g_5^4}{(4\pi)^4}V_G&\sum_{i=1}^{3}\Big(
	Y_{d}[\tilde{U}_\Delta]_{i1}[\tilde{U}_\Delta^*]_{i2}\\
	&+8\tilde{Y}[\tilde{U}_\Delta]_{i1}[\tilde{U}_\Delta^*]_{i3}
		\Big)I_{3}\Big(\frac{m^2_{\Delta_i}}{m_X^2}\Big)\,,
	\end{split}
\end{equation}
where the twiddles denote quantities in the rotated basis; note that the corresponding combination of $Y_{10}$ and $Y_{10}'$ denoted by $\tilde{Y}$ still remains free. However, since $|Y_d| \lesssim 10^{-2}$, the first term in~\eqref{eq:neutrino_formula_Yd} is unlikely to not play any significant role and in the remainder of this Appendix we shall stick to the second term therein and use \eqref{eq:neutrino_formula_Yd} in a simplified form
\begin{equation}\label{eq:neutrino_formula_simple}
        \begin{split}
		M_\nu^M\approx\frac{6g_5^4}{(4\pi)^4}V_G&\sum_{i=1}^{3}
        \Big(
		8\tilde{Y}[\tilde{U}_\Delta]_{i1}[\tilde{U}_\Delta^*]_{i3}
		\Big)
        I_{3}\Big(\frac{m^2_{\Delta_i}}{m_X^2}\Big)\,,
	\end{split}
\end{equation}
with the sole constraint on it arising from the requirement of perturbativity imposed on $\tilde{Y}$.

Without diving into unnecessary details (and following the same strategy as in previous studies) this constraint shall assume a somewhat na\"\i ve but conservative form of a limit on the absolute size of every element of the effective Yukawa coupling matrix governing the Feynman diagrams in Fig.~\ref{fig:graphs} in the form
\begin{equation}\label{limit 4pi}
	\left| 8\tilde{Y}\right| \le 4\pi\,.
\end{equation}
Moreover, since the weighted\footnote{Note that the relevant weights are elements of a unitary matrix.} sum of the loop integrals corresponding to the three triplet mass eigenstates obeys 
\begin{equation}
\left|
\sum_{i=1}^3[\tilde{U}_\Delta]_{i1}[\tilde{U}_\Delta^*]_{i3}I_{3}\left(\frac{m^2_{\Delta_i}}{m_X^2}\right)
\right| \le 3\,,
\end{equation}
cf.~\cite{Harries:2018tld}, one finally arrives at
\begin{equation}\label{eq:inequality MM}	\left| M_\nu^M\right|\le \frac{18g_5^4}{(4\pi)^3}V_G\,.
\end{equation}
\section{Efficient parameter space scanning \label{app:generator}}
In this Appendix we summarize the main analytical insights which  made it possible to significantly improve the efficiency of the numerical scans over the multi-dimensional parameter space of the model.   
\subsection{General implications of a lower bound on \texorpdfstring{$M_1$}{M1}\label{sec:M1limit}}
As it was argued in Section~\ref{sect:parameter_space}, barring severely fine-tuned scenarios (like resonant leptogenesis), in the current scenario it makes little sense to consider situations with $M_1$ below about $10^{7}$~GeV. Let us further discuss the consequences of such a constraint here.

Note that $M_1$ is the smallest element on the diagonal of $D_\nu^M$ from the Takagi decomposition of $M_\nu^M$ of~\eqref{seesaw_diagonal_form}, i.e., it is the square-root of the smallest eigenvalue of the matrix $M_\nu^M(M_\nu^M)^\dagger$. At the same time, it can be also viewed as the inverse value of the square-root of the largest eigenvalue of the matrix $|A|^2\equiv AA^\dagger$, where $A=(M_\nu^M)^{-1}$. Since the sum of the eigenvalues of $|A|^2$ is equal to its trace, one has
\begin{equation}
\label{ineqM1_1}
	\frac{3}{M_1^2}\ge\frac{1}{M_1^2}+\frac{1}{M_2^2}+\frac{1}{M_3^2}=\Tr A A^\dagger=\sum_{ij}\left| A_{ij}\right|^2\ge \left| A_{ij}\right|^2.
\end{equation}
for all $i$ and $j$.
Substituting for $A$ from~\eqref{original mLL} and~\eqref{diagonalization mLL} the inequality~\eqref{ineqM1_1} is brought to a simple form
\begin{equation}
	\sqrt{3} M_{1}^{-1}\geq\left| (\hat{D}_u^{-1} U_\nu^\dagger D_\nu U_\nu^* \hat{D}_u^{-1})_{ij} \right|=\left| (\hat{D}_u^{-1} m_{LL} \hat{D}_u^{-1})_{ij} \right|.
\end{equation}
which can be further recast as
\begin{equation}
\label{M1limits6}
    \left| (m_{LL})_{ij}\right|\le\N_{ij}\qquad \forall i,j
\end{equation}
where $m_{LL}=U_\nu^\dagger D_\nu U_\nu^*$ is the light neutrino mass matrix with elements in the sub-eV domain and
\begin{equation}
	  \N_{ij}=\sqrt{3}\,\frac{(\hat{D}_u)_i (\hat{D}_u)_j}{M_1}\,,
\end{equation}
is a symmetric matrix of the form
\begin{equation}
		\N\sim  \left(\frac{10^7\GeV}{M_1}\right)\!\!
	\begin{pmatrix}
		4.0\times 10^{-5}& 1.9\times 10^{-2}& 6.0\\
		. & 9.1 & 2.9\times 10^{3} \\
		. & . & 9.0\times 10^{5}
	\end{pmatrix}\!\!\eV.
\end{equation}
It is clear that 
\begin{equation}\label{leptogenesis 11 condition}
\left|\sum_k\left(U_\nu^*\right)_{k1}^2 \hat{m}_k\right|\le \N_{11}.
\end{equation}
is the most stringent out of the 6 independent bounds of Eq.~\eqref{M1limits6} irrespective of the actual value of $M_1$; hence, in what follows we shall predominantly focus on the constraints stemming from~\eqref{leptogenesis 11 condition}.
\subsection{Admissible shapes of the $U_\nu$ matrix}
\subsubsection{Restrictions on $U_\nu$ from the lower limit on $M_1$\label{sect:UnufromlimitonM1}}
For the normal neutrino hierarchy 
the general triangle inequalities 
\begin{align}
|z_1+z_2+z_3|\geq |z_2|-|z_1|-|z_3|\\ |z_1+z_2+z_3|\geq |z_3|-|z_1|-|z_2|  
\end{align}
applied to the triplet of complex numbers $z_k\equiv\left(U_\nu^*\right)_{k1}^2 \hat{m}_k$ inside the absolute value on the LHS of~\eqref{leptogenesis 11 condition} yield
\begin{align}\label{U21limit}
&\left|\left(U_\nu\right)_{21}\right|^2\le \frac{\hat{m}_3+\N_{11}}{\hat{m}_2+\hat{m}_3}\,,\\
\label{U31limit}
&\left|\left(U_\nu\right)_{31}\right|^2\le \frac{\hat{m}_2+\N_{11}}{\hat{m}_2+\hat{m}_3}\,.
\end{align}
Clearly, if the light neutrino spectrum is close to degenerate (i.e., for $m_2 \sim m_3 \gtrsim 5\times 10^{-2}$~eV), one obtains
\begin{align}
\label{U21limitN}
&\left|\left(U_\nu\right)_{21}\right|^2\lesssim 0.6\,,\\
\label{U31limitN}
&\left|\left(U_\nu\right)_{31}\right|^2\lesssim 0.4\,,
\end{align}
for any $M_1>10^7$~GeV.  
On the other hand, for hierarchical spectrum with $m_1\lesssim 10^{-3}\eV$
\eqref{U21limitN} is essentially gone whilst~\eqref{U31limitN} is strengthened to about 
\begin{equation}
    \left|\left(U_\nu\right)_{31}\right|^2\lesssim 0.15\,.
\end{equation}

It is worth noting that in either case these limits translate (cf. Eq.~\eqref{PMNS app}) into an interesting upper bound on one of the key factors driving proton decay in the current scenario, namely, 
\begin{equation}\label{Ul21_limit}
\left| U_\ell^L \right|^2_{21}\lesssim 0.8\,,
\end{equation}
see Section~\ref{sect:p-decay} and Fig.~\ref{fig:p-decay}.

For the inverted neutrino hierarchy
the same triangle inequalities provide
\begin{align}
\left|\left(U_\nu\right)_{11}\right|^2&\le \frac{\hat{m}_2+\N_{11}}{\hat{m}_2+\hat{m}_1}\,,\\
\left|\left(U_\nu\right)_{21}\right|^2&\le \frac{\hat{m}_1+\N_{11}}{\hat{m}_2+\hat{m}_1}\,.
\end{align}
which again lead to the bounds similar to~\eqref{U21limitN} and~\eqref{U31limitN}, just with permuted indices
\begin{align}
\label{U11limitNIH}
&\left|\left(U_\nu\right)_{11}\right|^2\lesssim 0.5\,,\\
\label{U21limitNIH}
&\left|\left(U_\nu\right)_{21}\right|^2\lesssim 0.5\,.
\end{align}
Since, however, $m_1$ and $m_2$ are always relatively large and close to each other in the IH case, these limits are quite insensitive to the specific shape of the light neutrino spectrum.

\subsubsection{Restrictions on $U_\nu$ from perturbativity\label{app:perturbativity_further}}
For very light lightest active neutrino, the bounds of the previous subsection can be complemented by the constraints from perturbativity discussed in brief in Appendix~\ref{app:perturbativity}. Indeed, the general limit~\eqref{eq:inequality MM} can be recast as 
\begin{equation}\label{eq:Inequalities of Perturbativity}
	\left| (\hat{D}_u U_\nu^T D_\nu^{-1} U_\nu \hat{D}_u)_{ij} \right| \le \frac{18\sqrt{2}g_5^3m_X}{(4\pi)^3}\equiv K\qquad \forall i,j
\end{equation}
by the virtue of the seesaw formula~\eqref{seesaw_diagonal_form}. Note that in terms of 
$m_{LL}^{-1} =U_\nu^T D_\nu^{-1} U_\nu
$
it assumes a particularly simple form 
\begin{equation}\label{eq:uznevim1}
     \left| (m_{LL}^{-1})_{ij}\right|\le K(\hat{D}_u)^{-1}_{i}(\hat{D}_u)^{-1}_{j} 
\end{equation}
 which, compared to inequality~\eqref{M1limits6}, makes the complementarity of the two sets of constraints evident.
 
Plugging in up-type quark masses (see Appendix~\ref{app:inputs}) the constraints of~\eqref{eq:uznevim1} can be recast as:
\begin{align}
\label{mLLconstraintN}
	&\left| \left(U_\nu^T\left(\frac{ D_\nu}{\rm eV}\right)^{\!-1}\!\!U_\nu\right)_{ij}\right|
 \leq
 \left(\frac{g_5}{0.5}\right)^3\left (\frac{m_X}{10^{17}\GeV}\right)\times\\
 &
 \qquad\times
\left(\begin{array}{ccc}
		7.0\times 10^{11}& 1.5\times 10^{9}& 4.6\times 10^{6}\\
		. & 3.0\times 10^{6} & 9.6\times 10^{3}\\
		.& . & 31
	\end{array}\right)_{\!\!ij}
\nonumber
\end{align}
With this at hand, several observations can readily be made:
\begin{enumerate}
   \item
    Concerning the general  decomposition of $U_\nu$ as in Eq.~\eqref{Unudecomposition}
    the inequalities~\eqref{mLLconstraintN} provide nontrivial constraits only on the $P_U^L$ and $U'_\nu$ matrices therein, and, hence, on the two phases $\beta_1$, $\beta_2$  of~\eqref{def:phases L}, and the three angles $\theta_{ij}$ and the $\delta$ phase parametrizing $U'_\nu$. The specific form of such bounds, however, strongly depends on $m_0$. 
    \item For $m_0\gtrsim 2\times 10^{-1}$~eV \eqref{mLLconstraintN} provide no constraints on the shape of the $U_\nu$ matrix whatsoever.
    \item The smaller $m_0$, the more specific the shape of $U_\nu$ has to be in order to conform all the constraints in~\eqref{mLLconstraintN}. Typically, the most important of these would be the one corresponding to $i=j=3$. This, however, does not need to be the case for really tiny $m_0$'s if $U_\nu$ assumes one of the special shapes which would keep the corresponding large element in $(D_\nu/{\rm eV})^{-1}$ entirely out of the second and third rows of $m_{LL}^{-1}$ on the LHS of~\eqref{mLLconstraintN}. At the extreme, taking $|(U_\nu)_{11}|=1$ in the normal hierarchy case (or $|(U_\nu)_{31}|=1$ for inverted hierarchy), $(\hat{m}
_0/{\rm eV})^{-1}$ would be fully contained at the $11$ position where it would encounter the weakest limit, i.e., the largest number on the RHS of~\eqref{mLLconstraintN}; this, in turn, would provide a universal absolute lower bound on $m_0$ in the current scenario in the form
\begin{equation}
\label{lower limit on m1}
	\quad \frac{m_0}{\rm eV}\gtrsim\left(\frac{0.5}{g_5}\right)^3\left (\frac{10^{17}\GeV}{m_X}\right)
	\times 1.4\times 10^{-12}\,.    
\end{equation}
\item 
For $m_0$ in the lower half between the no-constraint limit of $0.2$~eV and the absolute minimum of Eq.~\eqref{lower limit on m1} the $U_\nu$ matrices can be shown to assume the following approximate textures 
\begin{align}
\qquad&U_\nu\approx\begin{pmatrix}1& \epsilon_{12}\,\,\epsilon_{13}\\ {. \atop .} & V\end{pmatrix}\quad\text{with}\quad |\epsilon_{1i}|\sim \frac{\hat{m}_1 K}{(\hat{D}_u)_i^2}\quad\text{for NH}\,,\nonumber\\
\qquad&
U_\nu\approx\begin{pmatrix}{.\atop.} & V\\  1&\epsilon_{32}\,\,\epsilon_{33}\end{pmatrix}
\quad\text{with}\quad |\epsilon_{3i}|\sim \frac{\hat{m}_3 K}{(\hat{D}_u)_i^2}\quad\text{for IH}\,,\nonumber
\end{align}
where $V$ denote $2\times 2$ matrices which are unitary ($V^\dagger V=1$) up to the ${\cal O}(\epsilon_{ij})$ factors, and the small dotted entries are determined from the overall unitarity of $U_\nu$. Note that such a behaviour of $U_\nu$ in the NH case has been indicated already in~\cite{Rodriguez:2013rma}, see Fig.~3 therein. 
\end{enumerate}
\subsection{Efficient parameter space scanning technique}

\subsubsection{Sampling of the $U_\nu$ matrix}
For the sake of the numerical analysis, one has to efficiently generate a large sample of $U_\nu$ matrices conforming the previously discussed constraints, in particular~\eqref{leptogenesis 11 condition}. In the notation of Appendix~\ref{App:Basis change} (namely, Eq.~\eqref{Unudecomposition}) this amounts to choosing the two phases in the $P_U^L$ matrix of Eq.~\eqref{def:phases L} 
together with the three angles  $\theta_{ij}$ and one phase $\delta$ governing $U_\nu'$.

The key condition~\eqref{leptogenesis 11 condition} turns out to assume a particularly simple form
\begin{align}
\label{tangos}
&\N_{11}\ge \left|\hat{m}_3 \sin^2\phi_{13}\right.
\\&\left.+\cos^2\phi_{13}\,e^{-i(2\delta+\beta_2})\left(\hat{m}_1 \cos^2\phi_{12}+\hat{m}_2e^{i(\beta_2-\beta_1)}\sin^2\phi_{12}\right)\right|\nonumber
\end{align}
for $U'_\nu$ parametrized as
\begin{equation}
\label{cleverparametrization}
U'_\nu= o_{12}(\phi_{12}) o_{13}(\phi_{13},\delta) o_{23}(\phi_{23}),
\end{equation}
where $o_{ij}(\phi)$ are real orthogonal matrices corresponding to rotations in the $ij$ plane (with $-\sin \phi$ on the $ji$ position) and
\begin{equation}\label{endef:Unu cplx}
	o_{13}(\phi,\delta)\equiv\begin{pmatrix}
        \cos\phi & 0 & \sin\phi\,e^{-i\delta} \\
        0 & 1 & 0 \\
        -\sin\phi\,e^{i\delta} & 0 & \cos\phi
 \end{pmatrix}. 
\end{equation}
For a fixed lightest neutrino mass (out of which the entire light spectrum can be reconstructed), and specific values of $\phi_{12}$ and the $\beta_{1,2}$ and $\delta$ phases, \eqref{tangos} represents a quadratic inequality for $\sin^2\phi_{13}$.
Note that in the parametrization~\eqref{cleverparametrization} $\phi_{23}$ entirely drops from~\eqref{tangos}! This feature, together with some further insights on how to choose the physical Majorana phases $\alpha_1$ and $\alpha_2$ (cf. Eq.~\eqref{PMNS app2}) described below provides the key to the efficiency of our sampling procedure. 

\subsubsection{Sampling of the $\alpha_{1,2}$ Majorana phases\label{app:generator_phases}}
\begin{figure}
	\centering\includegraphics[width=8.5cm]{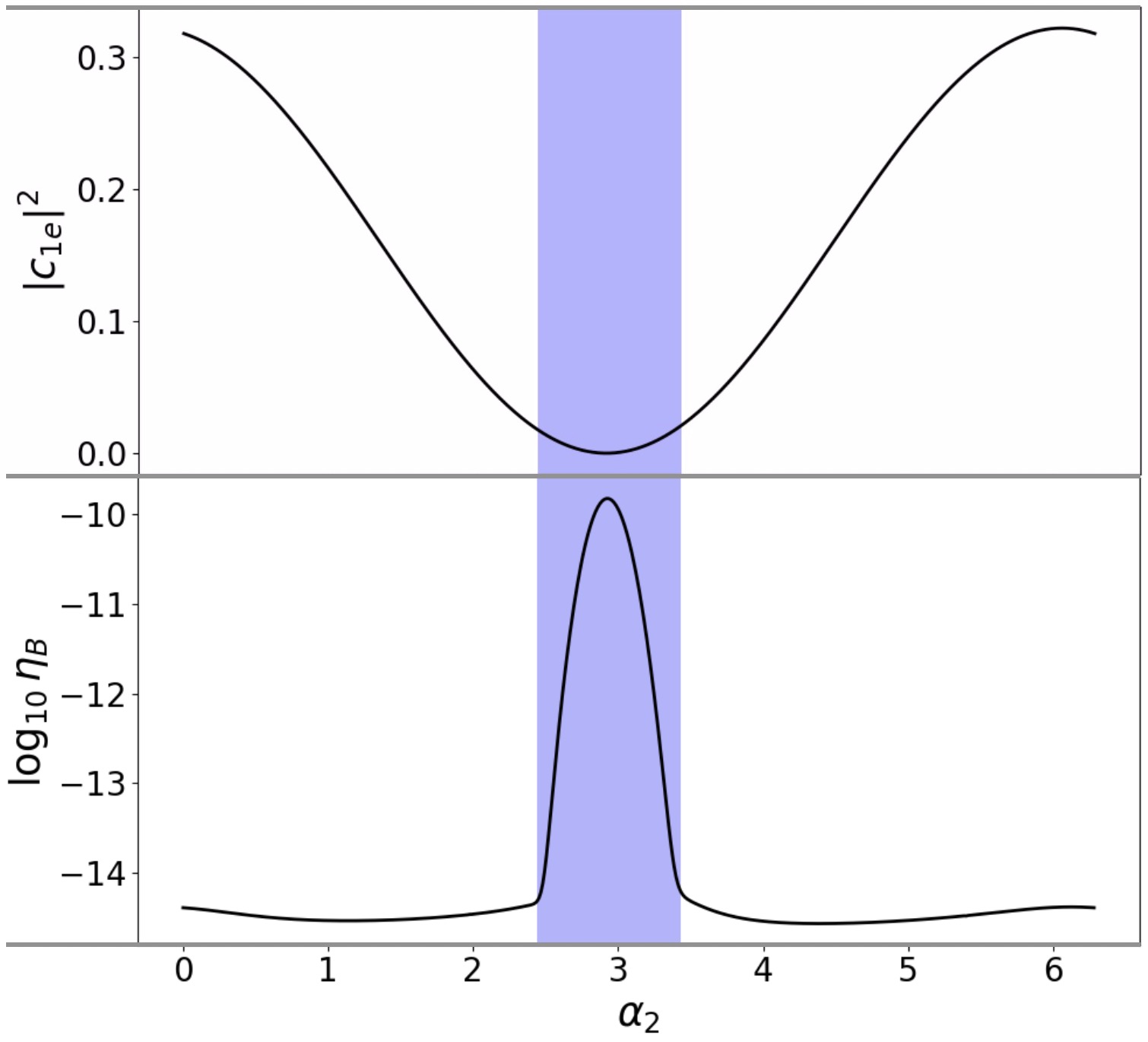}
	\caption{An illustration of the  behaviour of the  $c_{1\alpha}$ factor of Eq.~\eqref{eq:lepton_c} and the baryon asymmetry $\eta_B$ as a function of the $\alpha_2$ Majorana phase of~\eqref{def:phases L}
    for sample values of the other free parameters $\hat{m}_0=10^{-3}\eV$, $\sin\phi_{12}=0.239$, $\sin\phi_{13}=0.064$, $\sin\phi_{23}=0.551$, $\delta=6.27$, $\beta_1=-9.40$, $\beta_2=-6.27$ and $\alpha_1=1.36$, cf. Appendix~\ref{app:generator_phases}. Note that the enhancement of $\eta_B$ in the blue band area is behind the emergence of type-C Domain discussed in  Section~\ref{sect:Results} where the $N_1$ effects are generally suppressed. 
    \label{fig:etab_vs_majorana}} 
\end{figure}

Concerning the Majorana phases $\alpha_{1,2}$, there is nothing wrong in principle about sampling them randomly with a uniform distribution over the entire $\langle 0,2\pi)$ domain.
However, in attempt to boost the efficiency of finding large-$\eta_B$ points, it is highly beneficial to focus on special regions where, e.g., decoherence effects are small due to the suppression of $c_{1\alpha}$ or $c_{2\alpha}$ driving the charged current interactions of specific neutrino flavors. 
The same strategy can also help with staying out of the strong washout regime, as the washout factors in the context of Boltzmann equations~\eqref{Boltzmann_DE} typically appear in the $k_{i}c_{i\alpha}c_{i\beta}^{\star}$ combinations. This is illustrated in Fig.~\ref{fig:etab_vs_majorana} where the typical behaviour of $c_{1e}$ and $\eta_{B}$ as functions of one of the undelying phases is depicted (with all other parameters fixed).
Hence, the scans of Section~\ref{sect:Results} are predominantly focusing on the regions encompassing the minima  of $c_{i\alpha}$'s with less attention paid to points outside these domains.


\section{Minimal washout regions\label{app:minimal_washout}}
In this appendix we provide a simple analytic argument to justify the emergence of bands of accidentally small washout in the $M_1-M_2$ plane like the one encountered in the right panel of Fig.~\ref{fig:Domains_BC_kepsilon}. 
\subsection{The $k_2$ suppression band of Fig.~\ref{fig:Domains_BC_kepsilon} and $M_2$}
Given relation~\eqref{washout_ki}, the the small-washout regions are generally expected to correspond to the situation in which the $\tilde U$ matrix assumes a shape ensuring the total elimination of the $m_t^2$-proportional contribution therein, i.e., 
\begin{equation}
\label{key1}
    \tilde U_{23}=0\,. 
\end{equation}
In order to translate this condition into a specific constraint on $M_2$ (suggested by the shape of the suppression region  observed in Fig.~\ref{fig:Domains_BC_kepsilon}) it is convenient to rewrite the seesaw formula~\eqref{seesaw_diagonal_form} in the form
\begin{equation}
\mathbb{1}=- (D_\nu^M)^{-1/2}\tilde{U}^*\hat{D}_u U_\nu^T D_\nu^{-1} U_\nu \hat{D}_u \tilde{U}^\dagger (D_\nu^M)^{-1/2}
\end{equation}
which implies that 
\begin{equation}\label{Odef}
i(D_\nu^M)^{-1/2}\tilde{U}^*\hat{D}_u U_\nu^T (D_\nu)^{-1/2}=O
\end{equation}
must be a complex orthogonal matrix.
This comes very handy because the $k_2$ washout factor of our interest then obeys
\begin{equation}
\label{key2}
k_2=\frac{1}{m_*}({O} D_\nu{O}^\dagger)_{22}\,
\end{equation}
following from~\eqref{Yukawa_from_U}. 
Let us note that this is a universal form of $k_2$ in any model implementing type-I seesaw. Its added value in the current scenario is that it admits to relate the value of $M_2$ in the minimal washout regions of the parameter space to the quark masses via \begin{equation}
\label{key3}
M_2\sim\sum_{i=1}^3 m_i^{-1}\left|(\tilde{U}^*\hat{D}_u U_\nu^T)_{2i}\right|^2\,,
\end{equation}
which is a direct consequence of Eq.~\eqref{Odef} and the fact that $k_2$ of Eq.~\eqref{key2} is minimized for
\begin{equation}\label{minimum}
\sum_{i=1}^3 |O_{2i}|^2\sim 1\,.
\end{equation}
Inserting~\eqref{key1} into~\eqref{key3}  one obtains
\begin{equation}
M_2\sim\sum_{i=1}^3 m_i^{-1}\left|\sum_{j=1,2}(\tilde{U}^*)_{2j}(\hat{D}_u)_j (U_\nu^T)_{ji}\right|^2
\end{equation}
which, up to the singular $\tilde{U}_{22}=0$ case, always picks up a factor of $m_c^2$ from $(\hat{D}_u)_j^2$, hence justifying the emergence of small washout band at the position characterized by  Eq.~\eqref{eq:stripM2}. 

Note that the $\tilde{U}_{22}= 0$ case is discarded by the fact that in combination with~\eqref{key1} one would have $|\tilde U_{21}|=1$ which, by virtue of the seesaw formula~\eqref{seesaw_diagonal_form}, implies also $|(U_\nu)_{21}|=1$; this, however, is incompatible with~\eqref{U21limitN} for $\hat{m}_1=10^{-1.4}$~eV of Fig.~\ref{fig:Domains_BC_kepsilon} (but, as suggested by the widening of the green area towards the left side of the right panel therein, may become more viable for smaller $\hat{m}_1$ or a weaker limit on $M_1$, see Appendix~\ref{sec:M1limit}). 

\bibliography{bibliography}

\end{document}